\documentclass[useAMS,usenatbib]{mn2e}

\input{psfig.sty}
\usepackage{graphicx}
\usepackage{epsfig}
\usepackage{amssymb}
\usepackage{amsmath}
\usepackage{amsfonts}
\usepackage{txfonts}
\usepackage{multirow}
\usepackage{afterpage}
\usepackage{color}

\title[KiDS-450 Shear $\times$ {\it Planck} Lensing]{KiDS-450: Tomographic Cross-Correlation of Galaxy Shear with {\it Planck} Lensing}
\author[Harnois-D\'{e}raps, Tr\"oster, Chisari et al.]{Joachim Harnois-D\'{e}raps$^{1}$\thanks{E-mail: jharno@roe.ac.uk}, 
Tilman Tr\"oster$^{2}$, {\color{black} Nora Elisa Chisari}$^{3}$, {\color{black} Catherine Heymans}$^{1}$, 
 \newauthor {\color{black} Ludovic van Waerbeke}$^{2}$, Marika Asgari$^{1}$, Maciej Bilicki$^{4,5}$, Ami Choi$^{1,6}$,  
 \newauthor Hendrik Hildebrandt$^7$,  Henk Hoekstra$^{4}$, Shahab Joudaki$^{8,9}$, Konrad Kuijken$^{4}$,  
 \newauthor Julian Merten$^{3}$, Lance Miller$^{3}$,  Naomi Robertson$^{3}$, Peter Schneider$^{7}$ \&   Massimo Viola$^{4}$.\\
$^{1}$Scottish Universities Physics Alliance, Institute for Astronomy, University of Edinburgh, Blackford Hill, Scotland, UK\\
$^{2}$Department of Physics and Astronomy, University of British Columbia, 6224 Agricultural Road, Vancouver, B.C., V6T 1Z1, Canada\\
$^{3}$Department of Physics, University of Oxford, Keble Road, Oxford OX1 3RH, UK\\
$^{4}$Leiden Observatory, Leiden University, P.O.Box 9513, 2300RA Leiden, The Netherlands\\
$^{5}$National Centre for Nuclear Research, Astrophysics Division, PO Box 447, PL-90-950 \L\'od\'z, Poland\\
$^{6}$Center for Cosmology and AstroParticle Physics, The Ohio State University, 191 West Woodruff Avenue, Columbus, OH 43210, USA\\
$^{7}$Argelander-Institut f\"ur Astronomie, Auf dem H\"ugel 71, 53121 Bonn, Germany\\
$^{8}$Centre for Astrophysics \& Supercomputing, Swinburne University of Technology, PO Box 218, Hawthorn, VIC 3122, Australia\\
$^{9}$ARC Centre of Excellence for All-sky Astrophysics (CAASTRO)\\
}

\begin{document}

\date{\today}

\pagerange{\pageref{firstpage}--\pageref{lastpage}} \pubyear{2016}

\maketitle

\label{firstpage}

\begin{abstract}
 We present the tomographic cross-correlation between galaxy lensing measured in the Kilo Degree Survey (KiDS-450) with overlapping lensing measurements of the cosmic microwave background (CMB), as detected by {\it Planck} 2015.   We compare our joint probe measurement to the theoretical expectation for a flat $\Lambda$CDM cosmology, assuming the best-fitting cosmological parameters from the KiDS-450 cosmic shear and {\it Planck} CMB analyses.    We find that our results are consistent within $1\sigma$ with the KiDS-450 cosmology, with an amplitude re-scaling parameter $A_{\rm KiDS} = 0.86 \pm 0.19$.  Adopting a {\it Planck} cosmology, we find our results are consistent within $2\sigma$, with $A_{\it Planck} = 0.68 \pm 0.15$.  We show that the agreement is improved in both cases when the contamination to the signal by intrinsic galaxy alignments is accounted for, increasing $A$ by $\sim 0.1$.  
This is the first tomographic analysis of the galaxy lensing -- CMB lensing cross-correlation signal, and is based on five photometric redshift bins.  
We use this measurement as an independent validation of the multiplicative shear calibration  and of the calibrated source redshift distribution
{\color{black} at high redshifts. }
We find that constraints on these two quantities are strongly correlated when obtained from this technique, which should therefore not be considered as a stand-alone competitive calibration tool. 
\end{abstract} 



\section{Introduction}
\label{sec:intro}


Recent observations of distinct cosmological probes are closing in on the few parameters that enter the standard model of cosmology \citep[see for example][and references therein]{2016A&A...594A..13P}.
Although there is clear evidence that the Universe is well described by the $\Lambda$CDM model,
some tensions are found between probes. For instance, the best fit cosmology
inferred from the observation of the Cosmic Microwave Background (CMB) in \citet{2016A&A...594A..13P} is in tension with some cosmic shear analyses \citep[]{2015MNRAS.451.2877M, KiDS450, 2016arXiv161004606J, 2017MNRAS.465.2033J},
 while both direct and strong lensing measurements of today's Hubble parameter $H_0$ are more than 3$\sigma$ away from the values inferred from the CMB \citep{2016JCAP...10..019B, HOLICOW}. 
{\color{black} At face value, these discrepancies} either point towards new physics {\color{black}\citep[for a recent example, see][]{2016arXiv161004606J}} or un-modelled systematics in any of those probes.
In this context, cross-correlation of different cosmic probes stands out as a unique tool, as many residual systematics that could contaminate
one data set are unlikely to correlate also with the other (e.g. `additive biases'). This type of measurement can therefore be exempt from un-modelled biases that might otherwise source the tension.
Another point of interest is that the systematic effects that do not fully cancel, for example `multiplicative biases' or the uncertainty on the photometric redshifts,
will often impact differently the cosmological parameters compared to the stand-alone probe, allowing for degeneracy breaking or improved calibration.

In this paper, we present the first tomographic cross-correlation measurement between CMB lensing and galaxy lensing,
based on the lensing map described in \citet{2015arXiv150201591P} and the  lensing data from the Kilo Degree Survey\footnote{KiDS: {\tt http://kids.strw.leidenuniv.nl}} presented in \citet[][KiDS hereafter]{KiDSLenS} and in the KiDS-450 cosmic shear analysis \citep{KiDS450}. 
The main advantage in this sort of measurement resides in it being free of 
uncertainty on galaxy bias, which otherwise dominates the error budget  in CMB lensing - galaxy position cross-correlations \citep{2015arXiv150203405O, 2015arXiv150705551G, 2016arXiv160207384B}.     
Over the last two years, the first lensing-lensing cross-correlations were used to measure $\sigma_8$ and $\Omega_{\rm m}$  \citep{2015PhRvD..91f2001H, 2015PhRvD..92f3517L},
by combining the CMB lensing data from the Atacama Cosmology Telescope  \citep{2014JCAP...04..014D} with the lensing data from the Canada-France-Hawaii Telescope Stripe 82  Survey \citep{2014RMxAC..44..202M}, and from the {\it Planck} lensing data and the Canada-France-Hawaii Telescope Lensing Survey \citep[][CFHTLenS hereafter]{2013MNRAS.433.2545E}.
 Since then, additional effects were found to contribute to the measurement, introducing extra complications in the interpretation of the signal.  
For instance, \citet{2014MNRAS.443L.119H} and \citet{TroxelIshak14} showed that  the measurement is likely to be contaminated 
by the intrinsic alignment of galaxies with the tidal field in which they live. 
{\color{black} At the same time, \citet{2016PhRvD..93j3508L} argued that this measurement could point instead to residual systematics in the multiplicative shear bias
 and proposed that the measurement itself could be used  to set constraints on the shear bias \citep[see also][]{2013arXiv1311.2338D}. 
Their results showed that large residuals are favoured,  despite the calibration accuracy claimed by the  analysis  of image  simulations tailored for the same survey} \citep{2013MNRAS.429.2858M}.
A recent analysis from \citet[][HD16 hereafter]{2016MNRAS.460..434H} suggested instead that the impact of catastrophic redshift outliers could be causing this apparent discrepancy, 
since these dominate the uncertainty in the modelling. They also showed that choices concerning the treatment of the masks can lead to biases in the measured signal,
and that the current estimators should therefore be  thoroughly calibrated on full light-cone mocks.

Although these pioneering  works were based on Fourier space cross-correlation techniques, more recent analyses presented results from configuration-space measurements, 
which are cleaner due to their insensitivity to masking.  
\citet[][K16 hereafter]{Kirk15} combined the CMB lensing maps from {\it Planck} and from the South Pole Telescope  \citep[][SPT]{2012ApJ...756..142V} with 
the Science Verification Data from the Dark Energy Survey\footnote{DES: {\tt www.darkenergysurvey.org}}. 
 Their measurement employed the {\small POLSPICE} numerical tool \citep{spice, polspice},
 which starts off with a pseudo-$C_{\ell}$ measurement that is converted into configuration space to deal with masks, then turned back into a Fourier space estimator. 
Soon after, HD16 showed consistency between pseudo-$C_{\ell}$ analyses and {\color{black} configuration space analyses of two-point correlation functions}, combining the {\it Planck} lensing maps with both CFHTLenS and the Red-sequence Cluster Lensing Survey \citep[][RCSLenS hereafter]{RCSLenS}. 
A similar configuration space estimator was recently used with {\it Planck} lensing and SDSS shear data \citep{2017MNRAS.464.2120S}, 
although the signal was subject to higher noise levels.

This paper directly builds on the K16 and HD16 analyses, utilising tools and methods described therein, but on a new suite of lensing data. 
The additional novelty here is that we perform the first tomographic CMB lensing -- galaxy lensing  cross-correlation analysis, where we split the galaxy sample into 5 redshift bins and 
examine the redshift evolution. This is made possible by the high quality of the KiDS photometric redshift  data, {\color{black}  by the extend of the spectroscopic matched sample,
and consequently by the precision achieved on the calibrated source redshift distribution \citep[see][for more details]{KiDS450}}.
It provides a new test of cosmology within the $\Lambda$CDM model, {\color{black} including} the redshift evolution of the growth of structure,
and also offers an opportunity to examine the tension between the KiDS and {\it Planck} cosmologies \citep[reported in][]{KiDS450}.
With the upcoming lensing surveys such as LSST\footnote{\tt www.lsst.org} and Euclid\footnote{\tt sci.esa.int/euclid}, 
it is forecasted that this type of cross-correlation analysis will be increasingly used to {\color{black} validate the data calibration} \citep{2016arXiv160701761S} 
and extract cosmological information in a manner that complements the cosmic shear and clustering data.

The basic theoretical background upon which we base our work is laid out in Section \ref{sec:th}. 
We then describe the data sets and our measurement strategies in Sections \ref{sec:data} and \ref{sec:measurement}, respectively.
Our cosmological  results  are presented in Section \ref{sec:results}. {\color{black}  We also describe therein a calibration analysis along the lines of  \citet{2016PhRvD..93j3508L}, 
this time focussing on high redshift galaxies for which the photometric redshifts and shear calibration are not well measured. 
Informed on cosmology from  lower redshift measurement, this self-calibration technique has the potential to constraint jointly the shear bias and the photo-$z$ distribution,
where other methods fail.} We conclude in Section \ref{sec:conclusion}.

The fiducial cosmology that we adopt in our analysis corresponds to the flat {WMAP9}+SN+BAO cosmology\footnote{Our fiducial cosmology consists of a flat $\Lambda$CDM universe in which the dark energy equation of state is set to $w=-1$.} \citep{2013ApJS..208...19H},
in which the matter density, the dark energy density, the baryonic density, the  amplitude of matter fluctuations, the Hubble parameter 
and the tilt of the matter power spectrum are described by  $(\Omega_{\rm m}, \Omega_{\Lambda}, \Omega_{\rm b}, \sigma_8, h, n_s) = (0.2905, 0.7095, 0.0473, 0.831, 0.6898, 0.969)$.
Aside from {\color{black} determining}  the overall amplitude of the theoretical signal from the $[\sigma_8 - \Omega_{\rm m}]$ pair, this choice has little impact on our analysis, 
{\color{black} as we later demonstrate}.
{\color{black} Future surveys will have the statistical power to constrain the complete cosmological set, but this is currently out of reach for a survey the size of KiDS-450.}
 {\color{black} We note that our fiducial cosmology is a convenient choice} that is consistent within $2\sigma$ with the {\it Planck}, KiDS-450, CFHTLenS,  and {WMAP}9+{ACT}+{SPT} analyses in the $[\sigma_8 - \Omega_{\rm m}]$ plane.
As such, it minimizes the impact of residual tension across data sets.

\section{Theoretical Background}
\label{sec:th}

Photons from the surface of last scattering are gravitationally lensed by large scale structures in the Universe before reaching the observer.
{\color{black} Similarly}, photons emitted by observed galaxies are lensed by the low redshift end of the same large scale structures.
The signal expected from a cross-correlation measurement between the two {\color{black} lenses} can be related to the fluctuations in their common foreground matter field,
{\color{black} more precisely by the}  matter power spectrum $P(k,z)$. The lensing signal is obtained from an extended first order Limber integration over the past light cone up to the horizon 
distance $\chi_{\rm H}$, weighted by geometrical factors $W^{i}(\chi)$, assuming a flat cosmology \citep{Limber1954, LimberExt, NonLimber}:
\begin{eqnarray}
C^{\kappa_{\rm CMB}\kappa_{\rm gal}}_{\ell} = \int_0^{\chi_{\rm H}} d\chi W^{{\rm CMB}}(\chi)W^{{\rm gal}}(\chi) P\left(\frac{\ell+1/2}{\chi};z\right).
\label{eq:limber_cross}
\end{eqnarray}
In the above expression, $\chi$ is the comoving distance from the observer, $\ell$ is the angular multipole, and $z$ is the redshift.
The lensing kernels are given by 
\begin{eqnarray}
W^{i}(\chi) = \frac{3 \Omega_{\rm m} H_{0}^{2}}{2 c^2} \chi g^i(\chi) (1 + z),
\label{eq:W-def}
\end{eqnarray}
with
\begin{eqnarray}
g^{\rm gal }(\chi) = \int_{\chi}^{\chi_H} {\rm d} \chi' \tilde{n}(\chi') \frac{\chi' - \chi}{\chi'} \mbox{\hspace{5mm}and} \nonumber 
\end{eqnarray}
\begin{eqnarray}
g^{\rm CMB}(\chi) = \left[1 - \frac{\chi}{\chi_*}\right]{\rm H}(\chi_* - \chi).
\label{eq:g-chi-def}
\end{eqnarray}
The constant $c$ is the speed of light in vacuum,  $\chi_*$ is the comoving distance to the surface of last scattering. 
The term $\tilde{n}(\chi)$ is related to the redshift distribution of the observed  galaxy sources, $n(z)$, by 
$\tilde{n}(\chi) = n(z) \mbox{d}z/\mbox{d}\chi$, which depends on the depth of the survey. 
The Heaviside function ${\rm H}(x)$ guarantees that no contribution comes from behind  {\color{black} the surface of last scattering} as the integration in equation \ref{eq:limber_cross} approaches  the horizon.

The angular cross-spectrum described by equation \ref{eq:limber_cross} is related to correlation functions in configuration space, in particular 
{\color{black} between the CMB lensing map and the tangential shear} \citep{1991ApJ...380....1M}: 
\begin{eqnarray}
	\xi^{\kappa_{\rm CMB}\gamma_{\rm t}}( \vartheta) = \frac{1}{2\pi} \int_0^\infty \mbox{d} \ell\ \ell\ C^{\kappa_{\rm CMB}\kappa_{\rm gal}}_\ell J_{2}(\ell  \vartheta),
	\label{eq:xi-prediction}
\end{eqnarray}
where, ${J_2}$ is the  Bessel function of the first kind of order 2, and the quantity $\vartheta$ represents the angular separation on the sky. 
Details about measurements of $C^{\kappa_{\rm CMB}\kappa_{\rm gal}}_{\ell}$ and the tangential shear $\gamma_{\rm t}$  -- relevant to equations   \ref{eq:limber_cross} and \ref{eq:xi-prediction} respectively -- are provided in Section \ref{sec:measurement}.

Our predictions are obtained from the {\small NICAEA}\footnote{\tt www.cosmostat.org/software/nicaea/} cosmological tool \citep{Nicaea}, assuming a non-linear power spectrum described by the \citet{Takahashi2012} revision of the {\small HALOFIT} model \citep{Smith2003}. 




\section{The Data Sets}
\label{sec:data}


\subsection{KiDS-450 lensing data}
\label{subsec:KiDSLenS_data}

\begin{figure}
\begin{center}
\hspace{-3mm}
\includegraphics[width=3.4in]{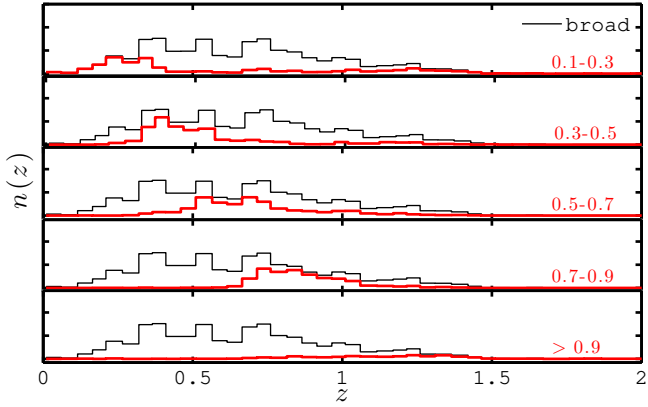}
\caption{Redshift distribution of the selected KiDS-450 sources in the tomographic bins (unnormalized), calibrated using the DIR method described in \citet{KiDS450}. 
The $n(z)$ of the broad $Z_B \in [0.1, 0.9]$ bin is shown in black in all panels for reference, while the $n(z)$ for the five tomographic bins are shown in red.
The mean redshift and effective number of galaxy in each tomographic bin are summarized in Table \ref{table:data}. }
\label{fig:nz}
\end{center}
\end{figure}

The KiDS-450 lensing data  that we use for our measurements are based on the third data release of {\color{black} dedicated KiDS} observations from the VLT Survey Telescope at Paranal, in Chile, 
and are described in \citet{KiDSLenS}, in \citet{KiDS450} and {\color{black} de Jong (2017 in prep.). These references describe the reduction and analysis pipelines leading to the shear catalogues, and present a rigorous and extensive set of systematic verifications}. Referring to these papers for more details, we summarise here the properties of the data that directly affect our measurement. 
 
Although the full area of the KiDS survey will consist of two large patches on the celestial equator and around the South Galactic Pole, the observing strategy was optimized to prioritize the coverage of  the GAMA fields \citep{2015MNRAS.452.2087L}. 
The footprint of the KiDS-450 data is consequently organized in five fields, G9, G12, G15, G23 and GS, covering a total of 449.7 deg$^2$
 {\color{black} While the  multiband imaging data are processed by Astro-WISE \citep{KiDS_DR2}},
the  lensing $r$-band data are processed by the  {\small THELI}  reduction method described in \citet{2013MNRAS.433.2545E}.
Shape measurements are determined using the self-calibrated {\it lens}fit  algorithm \citep[based on][]{2013MNRAS.429.2858M}
detailed in \citet{2016arXiv160605337F}. 

As described in \citet{KiDS450}, each galaxy is assigned a  photometric redshift probability distribution provided by the {\color{black} software} {\small BPZ} \citep{2000ApJ...536..571B}. The position of the maximum value of this distribution, labelled $Z_B$, serves only to divide the data into redshift bins. 
Inspired by the KiDS-450 cosmic shear measurement, we split the galaxy sample into 5 redshift bins: $Z_B \in [0.1, 0.3]$, $[0.3, 0.5]$, $[0.5, 0.7]$, $[0.7, 0.9]$ and $>0.9$. 
We also define a broad redshift bin by selecting all galaxies falling in the  range $Z_B \in [0.1, 0.9]$. 
The KiDS-450 cosmic shear measurement did not include the $Z_B>0.9$ bin because the photo-$z$ and the shear calibration {\color{black} were poorly constrained therein}. 
For this reason, we do not use this bin in our cosmological analysis either. Instead, we  estimate these calibration quantities directly from our measurement  in Section \ref{subsec:m}.

For each tomographic bin, the estimate of the redshift distribution of our galaxy samples, $n(z)$, is not obtained from the stacked {\small BPZ}-PDF, but from a magnitude--weighted scheme (in 4-dimensional $ugri$ magnitude space) of a spectroscopically matched sub-sample (the `weighted direct calibration' or `DIR' method from the KiDS-450 cosmic shear analysis, demonstrated to be the most precise method covering our redshift range).
Fig. \ref{fig:nz} shows these weighted $n(z)$ distributions, which enter the theoretical predictions through equation \ref{eq:limber_cross}, along with the effective number density per bin.
 In order to preserve the full description of the data in the high redshift tail, from where most of the signal originates, we do not fit the distributions with analytical functions,
as was done in previous work \citep[][K16, HD16]{2015PhRvD..91f2001H, 2015PhRvD..92f3517L}. Fitting functions tend to capture well the region where the $n(z)$ is maximal, 
however they attribute almost no weight to the (noisy) high redshift tail. This is of lesser importance in the galaxy lensing auto-correlation measurements, 
but becomes highly relevant for the CMB lensing cross-correlation.
Instead, we use the actual histograms in the calculation,
recalling that their apparent spikes are smoothed by the lensing kernels in equation \ref{eq:g-chi-def}. 
{\color{black} What is apparent from Fig. \ref{fig:nz}, and of importance for this analysis, is that the first tomographic bin has a long tail that significantly  overlaps 
with the CMB lensing kernel, more than the second tomographic bin. This feature is well captured by the mean redshift distributions, which are listed in  Table \ref{table:data}.}

\begin{table}
\caption{Summary of the data properties in the different tomographic bins.
The effective number of galaxy  assumes the estimation method of \citet{Heymans2012c}. }
\begin{center}
\begin{tabular}{cccc}
\hline
   $Z_B$  cut & $\bar{z}$ &  $n_{\rm eff}$ (gal/arcmin$^2$)  & $\sigma_{\epsilon}$  \\
\hline
\hline
\multirow{1}{*}{[0.1, 0.9]}
 & 0.72 &  7.54 & 0.28\\
\hline
\hline
\multirow{1}{*}{[0.1, 0.3]}
 & 0.75 &  2.23  & 0.29 \\
\hline
\multirow{1}{*}{[0.3, 0.5]}
 & 0.59 &  2.03 & 0.28 \\
\hline
\multirow{1}{*}{[0.5, 0.7]}
 & 0.72 &  1.81  & 0.27 \\
 \hline
\multirow{1}{*}{[0.7, 0.9]}
 & 0.87 &  1.49 & 0.28 \\
\hline
\multirow{1}{*}{$>$0.9}
& 1.27  &   0.90 & 0.33\\
 \hline
\end{tabular}
\end{center}
\label{table:data}
\end{table}%

Based on the quality of the ellipticity measurement, each galaxy is assigned  a {\it lens}fit weight $w$,  
plus a multiplicative shear calibration factor -- often referred to as the $m$-correction or the shear bias -- that is obtained from image simulations \citep{2016arXiv160605337F}. {\color{black} This calibration is accurate to better than 1\% for objects with $Z_B < 0.9$, but the precision quickly degrades at higher redshifts.  As recommended, we do not correct for shear bias in each galaxy, but instead compute the average correction for each tomographic bin (see equation \ref{eq:K}). 
In the fifth tomographic bin,  we expect to find residual biases in the  $m$-correction, but  apply it nevertheless, describing in Section \ref{subsec:m} how this correction can be self-calibrated. 
To be absolutely clear, we reiterate that we do not include this fifth bin in our main cosmological analysis.
The effective number density and the shape noise in each tomographic bin are also listed in  Table \ref{table:data}.

Following \citet{KiDS450},  we  apply a $c$-correction by subtracting the weighted mean ellipticity in each field and each tomographic bin,
but this has no impact on our analysis since this $c$ term does not correlate with the CMB lensing data. }

\subsection{{\it Planck} $\kappa_{\rm CMB}$ maps}
\label{subsec:CMB_data}

The CMB lensing data that enter our measurements are {\color{black} the $\kappa_{\rm CMB}$ map} obtained from the 2015 public data release\footnote{{\it Planck} lensing package: {\tt pla.esac.esa.int/pla/\#cosmology}}, thoroughly detailed in  \citet{2015arXiv150201591P}. The map making procedure is based on the quadratic estimator described in \citet{2002ApJ...574..566H},
generalized for a suite of multi-frequency temperature and polarization  maps. Frequencies are combined such as to remove foreground contamination, while other 
sources of secondary signal (mainly emissions from the galactic plane, from point sources and hot clusters) are masked in the CMB maps, prior to the reconstruction.  
If some of these are not fully removed from the lensing maps, they will create systematic effects in the $\kappa_{\rm CMB}$ map that show up differently
in the cross-correlation measurement compared to the auto-spectrum analysis. 
{\color{black} For example, there could be leakage in the CMB map coming from e.g. residual thermal Sunyaev-Zel'dovich signal that is  most likely located near massive clusters. 
These same clusters are highly efficient at lensing background galaxies, hence our cross-correlation measurement would be sensitive to this effect.
Indeed, the $\langle {\rm tSZ} \times \gamma_{\rm t}\rangle$, as recently measured in \citet{2016arXiv160807581H}, has a very large signal to noise
and could possibly be detected in a targeted analysis. Although it is difficult to assess the exact level of the tSZ signal in our  $\kappa_{\rm CMB}$ map,
the cleaning made possible from the  multi-frequency observations from {\it Planck} is thorough, reducing the residual contaminants to a very small fraction.
No quantitative evidence of such leakage has been reported as of yet, and we therefore ignore this in our analysis.  
}

Regions from the full sky lensing map that overlap with the five KiDS footprints are extracted, including a 4 degree extension to optimise the signal-to-noise ratio of the measurement (see HD16). 
The {\it Planck} release of lensing data also provides the analysis mask, which we apply to the $\kappa_{\rm CMB}$ map prior to carrying out our measurement\footnote{
This procedure does not entirely capture the masking analysis since the mask was applied on the temperature field, not on the  lensing map.
The reconstruction process inevitably leaks some of the masked regions into unmasked area, and vice versa. Applying this mask will therefore only remove the most problematic
regions. }.


\section{The measurements}
\label{sec:measurement}

\begin{figure*}
\begin{center}
\begin{tabular}{cc}
\hspace{-5mm}
\includegraphics[width=4.5in]{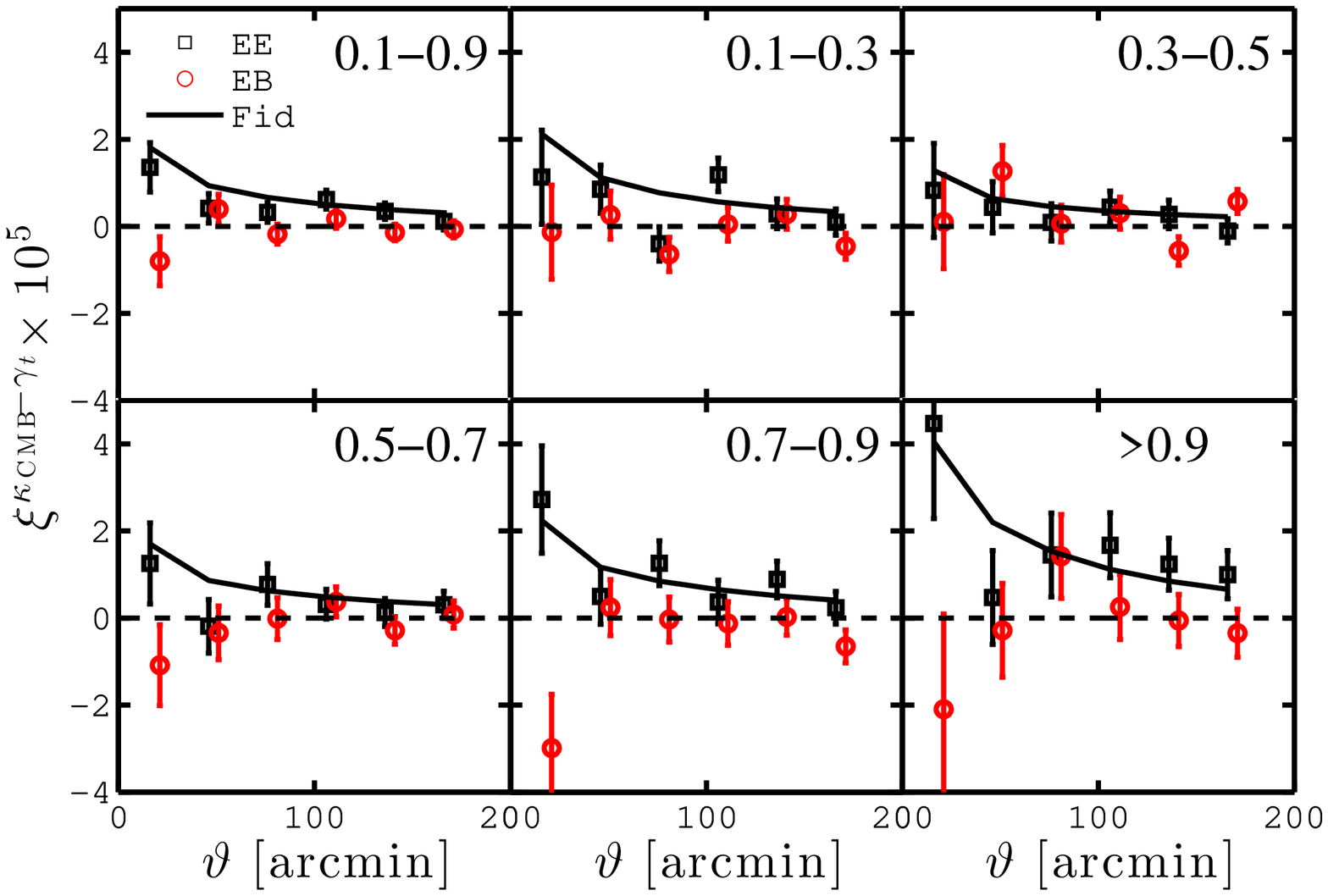}\\
\hspace{-7mm}
\includegraphics[width=4.45in]{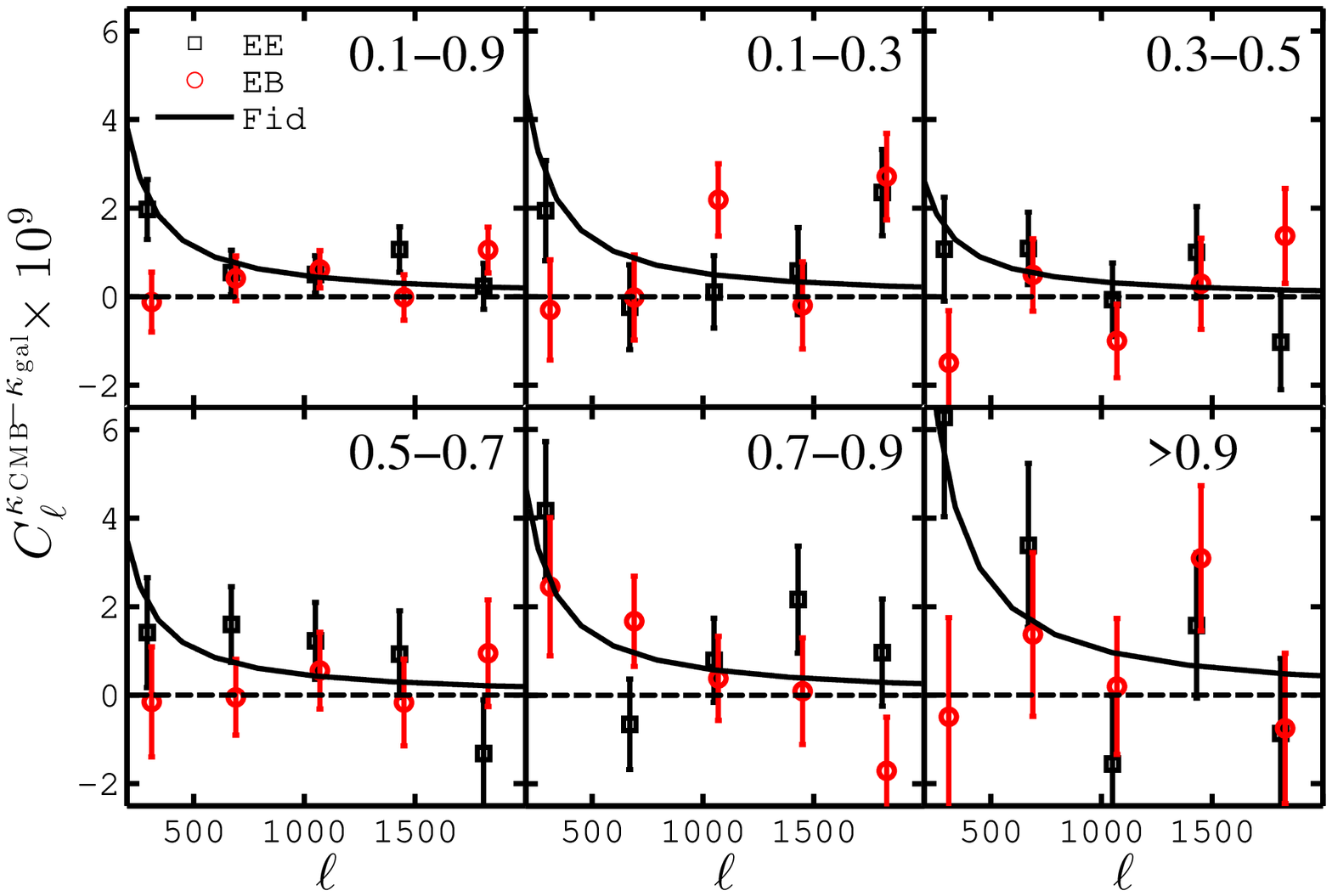}
\end{tabular}
\caption{Cross-correlation measurement between {\it Planck} 2015 $\kappa_{\rm CMB}$ maps and KiDS-450 lensing data.
The upper part presents results from the $\xi^{\kappa_{\rm CMB}\gamma_{\rm t}}$ estimator, 
while the lower part shows the estimation of $C_{\ell}^{\kappa_{\rm CMB} \kappa_{\rm gal}}$. 
Different panels show the results  in different tomographic bins, with predictions (solid curve) given  by equations \ref{eq:limber_cross} and  \ref{eq:xi-prediction} in our fiducial cosmology. 
The black squares show the signal, whereas the red circles present the EB null test described in Section \ref{subsec:null}, slightly shifted horizontally to improve the clarity in this figure. 
The error bars are computed from 100 CMB lensing simulations. } 
\label{fig:cmb_x_shear_KiDS_gamma_t}
\end{center}
\end{figure*}
 
 This section presents the cross-correlation measurements, which are performed with two independent estimators:
 $\xi^{\kappa_{\rm CMB}\gamma_{\rm t}}$ (equation \ref{eq:xi-prediction}) and the {\small POLSPICE} measurement of $C_{\ell}^{\kappa_{\rm CMB} \kappa_{\rm gal}}$ (equation \ref{eq:limber_cross}).
These were used and rigorously validated in previous work (see HD16, K16 and references therein for more details about these estimators).
 {\color{black} The reasons for conducting our analysis with two estimators instead of only one are two fold:  they do not probe the same scales, which makes them complementary,
 and, being completely independent, residual systematics could be identified through their different effect on these two statistics. }

\subsection{The $\xi^{\kappa_{\rm CMB}\gamma_{\rm t}}$ estimation}
\label{subsec:real_space}

The first estimator presented in this paper, $\xi^{\kappa_{\rm CMB}\gamma_{\rm t}}$, was recently introduced in HD16, and used later in \citet{2017MNRAS.464.2120S}.
It is a full configuration-space measurement that involves minimal manipulation of the data.
The calculation simply loops over each pixel of the $\kappa_{\rm CMB}$ maps and defines concentric annuli with different radii $\vartheta$, therein measuring the average tangential {\color{black} component of the} shear,
 $\gamma_{\rm t}$, from the KiDS galaxy shapes. For this reason, it is arguably the cleanest avenue  to perform such a cross-correlation measurement, even though there appears to be a 
limit to its accuracy at large angles in some cases due to the finite support of the observation window \citep{2013MNRAS.432.1544M}. That being said, it nevertheless bypasses a number of potential issues that are encountered with
other estimators (see HD16 for a discussion).
The  $\xi^{\kappa_{\rm CMB} \gamma_{\rm t}}$  estimator is given by:
\begin{eqnarray}
	\label{eq:shear-estimator}
	\xi^{\kappa_{\rm CMB}\gamma_{\rm t}}(\vartheta) = \frac{\sum_{ij} \kappa_{\rm CMB}^i e_{\rm t}^{ij} w^j \Delta_{ij}(\vartheta)}{\sum_{ij}  w^j \Delta_{ij}(\vartheta)}\frac{1}{1+K(\vartheta)} \,
\end{eqnarray}
where the sum first runs over the $\kappa_{\rm CMB}$ pixels `$i$', then over all galaxies `$j$' found in an annulus of radius $\vartheta$ and width $\Delta$, centered on the pixel $i$.
In this local coordinate system, $e_{\rm t}^{ij}$ is the tangential component of the {\it lens}fit ellipticity from the $j^{\rm th}$ galaxy relative to pixel $i$.
The exact binning scheme is described by $\Delta_{ij}(\vartheta)$, the binning operator:
\begin{eqnarray}
\Delta_{ij}(\vartheta)= \begin{cases} 
                                        1 {\rm , if } \left|{\boldsymbol \theta}_i - {\boldsymbol \theta}_j \right | < \vartheta \pm \frac{\Delta}{2} \\
                                        0 {\rm , otherwise} 
                                      \end{cases}
\label{eq:bin_operator}
\end{eqnarray}
where ${\boldsymbol \theta}_i$ and ${\boldsymbol \theta}_j$ are the observed positions of the pixel $i$ and galaxy $j$. 
{\color{black} Following HD16,} the bin width $\Delta$ is set to 30 arcmin, equally spanning the angular range [1, 181] arcmin with 6 data points. 
{\color{black} Larger angular scales  capture very little signal with the current level of statistical noise. 
We verified that our analysis results are independent of our choice of  binning scheme.} 
In  equation \ref{eq:shear-estimator}, $w^j$ is the {\it lens}fit weight of the galaxy $j$ and $K(\vartheta)$ corrects
for the shape multiplicative bias $m^j$ that must be applied to the lensing data  \citep{2016arXiv160605337F}:
 \begin{eqnarray}
	\frac{1}{1+K(\vartheta)} = \frac{\sum_{ij}  w^j \Delta_{ij}(\vartheta)}{\sum_{ij}  w^j (1+m^j)\Delta_{ij}(\vartheta)} \ .
	\label{eq:K}
\end{eqnarray}

The theoretical predictions for $\xi^{\kappa_{\rm CMB}\gamma_{\rm t}}$ are provided by equation \ref{eq:xi-prediction}.
We apply the same binning as with the data, averaging the continuous theory lines inside each angular bin.
We show in the upper panel of Fig. \ref{fig:cmb_x_shear_KiDS_gamma_t} the  measurements in all tomographic bins, 
compared to theoretical predictions given by our fiducial WMAP9+BAO+SN cosmology. 
The estimation of our error bars is described in Section \ref{subsec:error}.

We also project the galaxy shape components onto $e_{\times}$, which is rotated by 45 degrees compared to $e_{\rm t}$.
This effectively constitutes a nulling operation that can inform us of systematic leakage, in analogy to the EB test performed in the context of cosmic shear.
For this reason, we loosely refer to EE and EB tests in this paper, when we are in fact comparing $\kappa_{\rm CMB} \times e_{\rm t}$ and $\kappa_{\rm CMB} \times e_{\times}$, respectively. 
We note that the past literature referred to such a EB measurement as the `B-mode test',
which can be misleading for the non-expert. Indeed, the proper B-mode test refers to the BB measurement in weak lensing analyses,
a non-lensing signal that can be caused by astrophysics and systematics. The EB signal test {\color{black} asserts} something more fundamental: 
 since B changes sign under parity, and E does not, a non-zero EB means a violation of the parity of the shear/ellipticity field \citep{2003A&A...408..829S}. 
 That is not expected from astrophysics, so could only come from a systematic effect that does not vanish under averaging.
Our EB measurement is shown with the red symbols in Fig. \ref{fig:cmb_x_shear_KiDS_gamma_t}.
We find by visual inspection that in most tomographic bins, these  seem closely centered on zero, but not in all cases.
To quantify the significance of this EB measurement, we estimate  the confidence at which these red points deviate from zero. 
We detail in Section \ref{subsec:null} how we carry out that test and show that  they are consistent with noise.


We have carried out an additional null test presented in HD16, which consists in rotating randomly the shapes of the galaxies before the measurement 
($\kappa_{\rm CMB} \times \mbox{random}$). {\color{black} This test is sensitive to the noise levels in the galaxy lensing data and hence affected by the shape noise $\sigma_{\epsilon}$ listed in Table \ref{table:data}. We find that the resulting signal is fully consistent with zero in all tomographic bins. }


\subsection{The $C_{\ell}^{\kappa_{\rm CMB} \kappa_{\rm gal}}$ estimation}
\label{subsubsec:PolSpice}

The second estimator uses  the  same data  as our $\xi^{\kappa_{\rm CMB}\gamma_{\rm t}}$ analysis, namely the $\kappa_{\rm CMB}$ map and the KiDS shear catalogues,
but requires additional operations on the data, including harmonic space transforms. This is accomplished with the {\small POLSPICE} numerical 
code \citep{spice,polspice} running in polarization mode, where the \{T, Q, U\} triplets are replaced by {\color{black} \{$\kappa_{\rm CMB}, 0, 0\}$ and $\{0, -e_1, e_2$\}}. 
The code first computes the pseudo-$C_{\ell}$ of the maps and of the masks, then transforms the results  into configuration space quantities,
that are finally combined and transformed back into Fourier space. The output of   {\small POLSPICE} is therefore an estimate of the cross-spectrum $C_{\ell}^{\kappa_{\rm CMB} \kappa_{\rm gal}}$. 
While {\small POLSPICE}  is frequently used for CMB analyses, it was applied for the first time in the context of CMB lensing $\times$ galaxy lensing by K16 and serves as a good comparison to the configuration estimator described in Section 
\ref{subsec:real_space}. One main advantage of this estimator is that in principle different $\ell$-bands are largely uncorrelated, which makes the covariance matrix almost diagonal and  hence  easier to estimate.

The {\small POLSPICE} measurement\footnote{{\color{black} {\small POLSPICE}  has adjustable internal parameters, and we use {\sc thetamax} = 60 deg., {\sc apodizesigma} = 60 deg. and {\sc nlmax}  = 3000.}} is presented in the lower panel of Fig. \ref{fig:cmb_x_shear_KiDS_gamma_t}, plotted against the theoretical predictions given by equation \ref{eq:limber_cross}.
The EB data points are directly obtained from the temperature/B-mode output provided by the polarization version of the code,  
 and are further discussed in Section \ref{subsec:null}.

Note that our choice of the $\gamma_{\rm t}$ and  {\small POLSPICE} estimators was motivated by our desire to avoid producing $\kappa_{\rm gal}$ maps in order to reduce the risks of errors and systematic biases
that can arise in the map making stage in the presence of  a mask as inhomogeneous as that of the KiDS-450 data.
These two estimators produce correlated measurements, but the scales they are probing differ. 
The $\gamma_{\rm t}$ estimator is accurate at the few percent level, as verified on full mock data in HD16, and the {\small POLSPICE} code has been thoroughly
verified and validated on the same mocks as well. We refer the reader to K16 and HD16 for details of these tests. 


\subsection{Covariance Estimation}
\label{subsec:error}

The $\kappa_{\rm CMB}$ map reconstructed by the {\it Planck} data is noise dominated  for most Fourier modes \citep{2015arXiv150201591P}.
It is only by combining the full sky temperature and polarization maps that the {\it Planck} Collaboration could achieve a lensing detection of $40\sigma$. 

 Since the noise $N_{\rm CMB}$ is  larger than the signal $\kappa_{\rm CMB}$ at every scale included  in our analysis (HD16), 
 we can evaluate the covariance matrix from  cross-correlation measurements between the 
 100 {\it Planck} simulated lensing maps (also provided in their 2015 public data release) and the tomographic KiDS data:
\begin{eqnarray}
{\rm Cov}^{\kappa_{\rm CMB} \kappa_{\rm gal}}_{\ell \ell'} \simeq \langle \Delta \hat{C}^{N_{\rm CMB} \kappa_{\rm gal}}_{\ell}  \Delta \hat{C}^{N_{\rm CMB} \kappa_{\rm gal}}_{\ell'} \rangle
\end{eqnarray}
and
\begin{eqnarray}
{\rm Cov}^{\kappa_{\rm CMB} \gamma_{\rm t}}_{\vartheta \vartheta'} \simeq \langle \Delta \hat{\xi}^{N_{\rm CMB} \gamma_{\rm t}}_{\vartheta}  \Delta \hat{\xi}^{N_{\rm CMB} \gamma_{\rm t}}_{\vartheta'} \rangle.
\label{eq:Cov_approx}
\end{eqnarray}
where the `hats' refer to measured quantities, $\Delta \hat{x} = \hat{x} - \bar{x}$, and the brackets represent the average over the 100 realizations. 
{\color{black} This method assumes that the covariance is completely dominated by the CMB lensing and neglects the contribution from the shear covariance.
This is justified by the fact that the signal from the  former is about an order of magnitude larger, and  hence completely drives the statistical uncertainty (HD16).}
The error bars shown in Fig. \ref{fig:cmb_x_shear_KiDS_gamma_t}  are obtained from these matrices (from the square root of the diagonals).
For each tomographic bin, the ${\rm Cov}^{\kappa_{\rm CMB} \kappa_{\rm gal}}_{\ell \ell'}$ matrix has 25 elements, whereas the  ${\rm Cov}^{\kappa_{\rm CMB} \gamma_{\rm t}}_{\vartheta \vartheta'}$
matrix has 36. The 100 realizations are enough to invert these matrices one at a time with a controllable level of noise bias,
and the numerical convergence on this inverse is guaranteed \citep{Lu2010}.  
{\color{black} 
  Note that this strategy fails to  capture the correlation between tomographic bins, which are not required by our cosmological analysis presented in Section \ref{subsec:cosmo_broad}.
 If needed in a future analysis, these could be estimated from full light-cone mock simulations.}

For both estimators, the covariance matrix is dominated by its diagonal, with most off-diagonal elements of the cross-correlation coefficient matrix being under $\pm10\%$.
Some elements reach larger values, $\pm$40\% correlation at the most, but these are isolated,  not common to all tomographic bins, 
and are consistent with being noise fluctuations, 
given that we are measuring many elements from `only' 100 simulations.
This partly explains why our cosmological results are not based on a joint tomographic analysis.
We keep the full matrices in the analysis, even though we could, in principle, include only the diagonal part in the   {\small POLSPICE} measurement.
Nevertheless, we have checked that our final results  are only negligibly modified if we use this approximation in the $\chi^2$ calculation, suggesting that one could reliably use a {\color{black} Gaussian approximation to the error estimation in this type of measurement (see equation 23 in HD16)}.

\section{Cosmological Inference}
\label{sec:results}

Given the relatively low signal-to-noise ratio of our measurement (Fig. \ref{fig:cmb_x_shear_KiDS_gamma_t}), we do not fit our signal for the six parameters $\Lambda$CDM cosmological model.
Instead, we follow the strategy adopted by earlier measurements: {\color{black}  we compare the measured signal to our fiducial cosmological predictions, treating the normalization as a free parameter `$A$'.
If the assumed fiducial cosmology is correct and in absence of other systematic effects, $A$ is expected to be consistent with unity. }
As discussed in previous studies, $A$ is affected by a number of {\color{black} effects} that can similarly modulate the overall amplitude of the signal.
Aside from its sensitivity to cosmology -- our primary science target -- this rescaling term will absorb contributions from residual systematic errors in the estimation of $n(z)$, 
from mis-modelling of the galaxy intrinsic alignments, 
from residual systematic bias in the shear multiplicative term $m$ (equation \ref{eq:K}), from  astrophysical phenomena such as massive neutrinos and/or baryonic feedback, 
and from residual systematics in the cross-correlation estimators themselves (K16 and HD16). 

In this section, we first  present our constraints on $A$; we then quantify how the different effects listed above can impact our measurements, 
and finally present our cosmological interpretation. {\color{black} Our primary results assume the fiducial WMAP9+BAO+SN cosmology, i.e. we first place constraints on $A_{\rm fid}$, 
however  we also report constraints on $A_{\rm KiDS}$ and $A_{Planck}$, obtained by assuming different baseline cosmologies.}

\subsection{Significance}
\label{subsec:A}

To measure $A$, we first compute the $\chi^2$ statistic:
\begin{eqnarray}
\chi^2 =  \Delta{\rm x}^{\rm T}\ \mbox{Cov}^{-1}\  \Delta{\rm x}
\label{eq:chi}
\end{eqnarray}
 with
 \begin{eqnarray}
  \Delta{\rm x} = \hat{\xi}^{\kappa_{\rm CMB}\gamma_{\rm t}} - A {\xi}^{\kappa_{\rm CMB}\gamma_{\rm t}} \mbox{ or }  \Delta{\rm x} = \hat{C}^{\kappa_{\rm CMB}\kappa_{\rm gal}} - A {C}^{\kappa_{\rm CMB}\kappa_{\rm gal}}
  \label{eq:chi2}
\end{eqnarray}
 for the configuration space and {\small POLSPICE} estimators, respectively. {\color{black} As before, quantities with `hats' are measured, and the predictions assume the fiducial cosmology, 
 unless stated otherwise}.
 The signal-to-noise (SNR) ratio is given by the likelihood ratio test, 
which measures the confidence at which we can reject the null hypothesis (i.e. that there is no signal, simply noise)
in favour of an alternative hypothesis described by our theoretical model with a single parameter $A$ \citep[see][for a recent derivation  in a similar context]{2016arXiv160807581H}. 
We can write SNR $= \sqrt{\chi^2_{\rm null} - \chi^2_{\rm min}}$, where $\chi^2_{\rm null}$ is computed by setting $A = 0$, 
and $\chi^2_{\rm min}$ corresponds to the best-fit value for $A$.
The error on $A$ is obtained by varying the value of $A$ until $ \chi^2_{A} - \chi^2_{\rm min} = 1$ \citep[see, e.g.][]{2003psa..book.....W}.

We include two additional statistical corrections to this calculation.
The first is a correction factor that multiplies the inverse covariance matrix, $\alpha = (N_{\rm sim}-N_{\rm bin}-2)/(N_{\rm sim}-1) = 0.94$, to account for biases inherent to matrix inversion in the presence of noise \citep{Hartlap2007a}. Here $N_{\rm bin}$ is the number of data bins (5 for $C^{\kappa_{\rm CMB}\kappa_{\rm gal}}$ and 6 for $\xi^{\kappa_{\rm CMB}\gamma_{\rm t}}$) and $N_{\rm sim}$ is the number of simulations (100) used in the covariance estimation. 
There exists an improved version of this calculation  based on assuming a  $t$-distribution {\color{black} in the likelihood}, 
however {\color{black} with our values of  $N_{\rm bin}$ and $N_{\rm sim}$}, the differences in the inverted matrix would be of order 10-20\%  \citep{2016MNRAS.456L.132S},
a correction on the error that we ignore given the relatively high level of noise in our measurement. 

The second correction was first used in HD16, and consists of an additional error on $A$ 
due to the propagated uncertainty coming from the noise in the covariance matrix \citep{2014MNRAS.442.2728T}.
This effectively maps  $\sigma_A \rightarrow  \sigma_A(1 + \epsilon/2)$, where $\epsilon = \sqrt{2/N_{\rm sim}+ 2(N_{\rm bin}/N_{\rm sim}^2)} = 0.145 $. 
These two correction factors are included  in the analysis.  
The results from our statistical investigation are reported in Table \ref{table:stat}, where we list $\chi^2_{\rm min}$, $\chi^2_{\rm null}$, SNR and $A$ 
for every tomographic bin. The theoretical predictions provide a good fit to the data given that for our degrees of freedom $\nu = N_{\rm bin}-1$, 
$\nu - \sqrt{2\nu} < \chi^2_{\rm min} < \nu + \sqrt{2\nu}$. In other words, all our measured $\chi^2$ fall within the expected 1$\sigma$ error.
 {\color{black} We also compute the $p$-value for all these $\chi^2$ measurements at the best-fitting $A$ in order to estimate  the confidence at which we can accept or reject the assumed model.
 Assuming Gaussian statistics, $p$-values smaller than 0.01 correspond to a 99\% confidence in the rejection of the model (the null hypothesis) by the data, and are considered `problematic'.
 Our measured $p$-values, also listed in Table \ref{table:stat}, are always larger than  0.12, meaning that the model provides a good fit to the data in all cases.}

\begin{table}
\caption{Summary of $\chi^2$,  SNR {\color{black} and $p$-values} obtained with the two different pipelines. 
The $C_{\ell}^{\kappa_{\rm CMB} \kappa_{\rm gal}}$ measurements have  4 degrees of freedom (5 $\ell$-bins - 1 free parameter),
whereas the configuration space counterpart $\xi^{\kappa_{\rm CMB} \gamma_{\rm t}}(\vartheta)$ has one more, with 6 $\vartheta$-bins.
$A_{\rm fid}$  is the best-fit amplitude that scales the theoretical signals in the fiducial cosmology, according to equation \ref{eq:chi2},
also shown in Fig. \ref{fig:A_tomo}.
 The numbers listed here include the covariance debiasing factor $\alpha$ and the extra error $\epsilon$ due to the noise in the covariance (see main text of Section \ref{subsec:A} for more details). }
\begin{center}
\begin{tabular}{clccccc}
\hline
    $Z_B$  &  Estimator  & $\chi^2_{\rm min}$ & $\chi^2_{\rm null}$ & SNR & \color{black} $p$-values & $A_{\rm fid}$  \\ 
\hline
\hline
\multirow{2}{*}{[0.1, 0.9]}
 &  $C_{\ell}^{\kappa_{\rm CMB} \kappa_{\rm gal}}$ & 2.80 & 18.21 & 3.93 & 0.53& 0.77 $\pm$ 0.19 \\
 &  $\xi^{\kappa_{\rm CMB}\gamma_{\rm t}}$ & 2.88 & 22.94 & 4.48 & 0.64 & 0.69 $\pm$ 0.15 \\
\hline
\hline
\multirow{2}{*}{[0.1, 0.3]}
 &  $C_{\ell}^{\kappa_{\rm CMB} \kappa_{\rm gal}}$ & 5.48 & 8.89 & 1.85 & 0.20 & 0.55 $\pm$ 0.30 \\
 &  $\xi^{\kappa_{\rm CMB}\gamma_{\rm t}}$ & 7.93 & 13.38 & 2.34 & 0.12 & 0.53 $\pm$ 0.24 \\
\hline
\multirow{2}{*}{[0.3, 0.5]}
  &  $C_{\ell}^{\kappa_{\rm CMB} \kappa_{\rm gal}}$ & 2.95 & 4.95 & 1.42 & 0.50 & 0.71 $\pm$ 0.51 \\
 &  $\xi^{\kappa_{\rm CMB}\gamma_{\rm t}}$ & 1.44 & 4.19 & 1.66 & 0.84 & 0.60 $\pm$ 0.37 \\
\hline
\multirow{2}{*}{[0.5, 0.7]}
 &  $C_{\ell}^{\kappa_{\rm CMB} \kappa_{\rm gal}}$ & 4.00 & 10.13 & 2.47 & 0.35 & 0.87 $\pm$ 0.35 \\
 &  $\xi^{\kappa_{\rm CMB}\gamma_{\rm t}}$ & 2.00 & 6.45 & 2.11 & 0.77 & 0.55 $\pm$ 0.26 \\
 \hline
\multirow{2}{*}{[0.7, 0.9]}
 &  $C_{\ell}^{\kappa_{\rm CMB} \kappa_{\rm gal}}$ & 5.12 & 10.04 & 2.22 & 0.23 & 0.79 $\pm$ 0.36 \\
 &  $\xi^{\kappa_{\rm CMB}\gamma_{\rm t}}$ & 2.78 & 15.41 & 3.55 & 0.65 & 1.02 $\pm$ 0.29 \\
\hline
\multirow{2}{*}{$>0.9$}
 &  $C_{\ell}^{\kappa_{\rm CMB} \kappa_{\rm gal}}$ & 4.70 & 12.92 & 2.87 & 0.26 & 0.83 $\pm$ 0.29 \\
 &  $\xi^{\kappa_{\rm CMB}\gamma_{\rm t}}$ & 4.68 & 22.64 & 4.24 & 0.38 & 0.95 $\pm$ 0.22 \\
 \hline
\end{tabular}
\end{center}
\label{table:stat}
\end{table}%

\begin{figure}
\begin{center}
\begin{tabular}{cc}
\hspace{-5mm}
\includegraphics[width=3.2in]{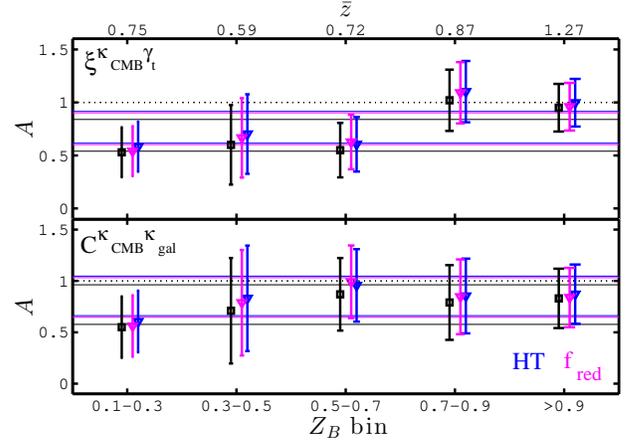}
\end{tabular}
\caption{Tomographic measurement of $A_{\rm fid}$, defined in equation \ref{eq:chi2}, assuming our fiducial cosmology. 
The two panels present results from the two cross-correlation estimators (labelled in the top left corner). 
Black symbols assume no IA, while color symbols include correction factors from two IA models ($f_{\rm red}$ in magenta and HT in blue, see Sec. \ref{subsec:IA}).
The horizontal solid lines of a given color enclose the $1\sigma$ region measured in the broad  $Z_B \in [0.1, 0.9]$ bin, 
while the dotted horizontal lines indicate  the fiducial values ($A_{\rm fid}$ = 1).
The mean source redshift in each bin is indicated at the top and summarized in Table \ref{table:data}.
{\color{black} The mean in the first bin is high because of the long tail, visible in Fig. \ref{fig:nz}. 
 The best fit values in different cosmologies are $A_{\rm fid} = 0.69\pm0.15$, $A_{\rm KiDS} = 0.86\pm0.19$ and $A_{Planck} = 0.68\pm0.15$.}
} 
\label{fig:A_tomo}
\end{center}
\end{figure}

\begin{figure}
\begin{center}
\begin{tabular}{cc}
\hspace{-5mm}
\includegraphics[width=3.2in]{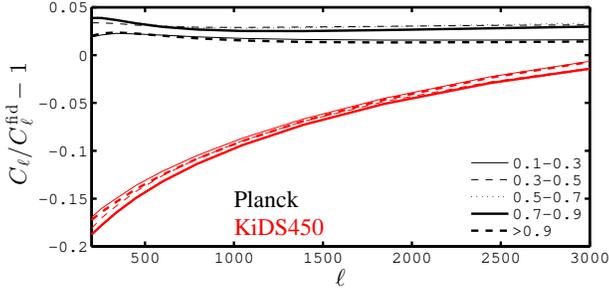}
\end{tabular}
\caption{Fractional effect on the signal when changing the fiducial cosmology  to {\it Planck} or KiDS-450.
Different symbols show the impact in different tomographic bins, relative to the fiducial predictions.
Current measurements are limited to $\ell<2000$.} 
\label{fig:frac_cosmo}
\end{center}
\end{figure}

These tomographic measurements  are re-grouped in Fig. \ref{fig:A_tomo}, where we compare the redshift evolution of $A$ for both estimators. 
We mark the $1\sigma$ region of the broad bin $n(z)$ with the solid horizontal lines, and see that all points overlap with this region within $1\sigma$.
This is an indication that the {\color{black} relative growth of structure between the tomographic bins} is consistent with the assumed $\Lambda$CDM model.
For the broad $n(z)$, the signal prefers an amplitude that is $\sim23-31$ percent lower than the fiducial cosmology, 
 i.e.  the $1\sigma$ region shown by the horizontal solid lines in Fig. \ref{fig:A_tomo} is offset from unity by that amount.
The main cosmological result that we quote from the $Z_B \in [0.1, 0.9]$ measurement is that of the $\gamma_{\rm t}$  estimator due to its higher SNR, 
as seen from comparing the top two rows of Table \ref{table:stat}.
 For our fiducial cosmology, we find:
\begin{eqnarray}
A_{\rm fid} = 0.69\pm0.15
\label{eq:A_fid}
\end{eqnarray}
%
Varying the cosmology to the best fit KiDS-450
and {\it Planck} cosmologies\footnote{\color{black} Fiducial $(\Omega_{\rm m}, \Omega_{\Lambda}, \Omega_{\rm b}, \sigma_8, h, n_s) = (0.29, 0.71, 0.047, 0.83, 0.69, 0.97)$.}$^,$\footnote{KiDS-450 $(\Omega_{\rm m}, \Omega_{\Lambda}, \Omega_{\rm b}, \sigma_8, h, n_s) = (0.25, 0.75, 0.047, 0.85, 0.75, 1.09)$.}$^,$\footnote{{\it Planck} $(\Omega_{\rm m}, \Omega_{\Lambda}, \Omega_{\rm b}, \sigma_8, h, n_s) = (0.32, 0.68, 0.049, 0.80, 0.67, 0.97)$.}, we obtain 
\begin{eqnarray}
A_{\rm KiDS} = 0.86\pm0.19 \mbox{\hspace{5mm} and \hspace{5mm} } A_{Planck} = 0.68\pm0.15
\end{eqnarray}
The relative impact of these different cosmologies on our signal is presented in Fig. \ref{fig:frac_cosmo}, where we see that the KiDS-450 cosmology mostly differ from the other two at large scales.
The signal from our fiducial cosmology agrees with that assuming the best-fit {\it Planck} cosmology  to better than 5\% in all tomographic bins.

{\color{black}  Fig. \ref{fig:A_tomo} demonstrates there are small but noticeable differences between the two estimators at fixed cosmology, especially at high redshift.
As mentioned in Section \ref{subsubsec:PolSpice},  the scales being probed are not identical, and therefore 
 some differences in the recovered values of $A$  are expected. 
 Nevertheless, within the current statistical accuracy, the two estimators are fully consistent with one another. }

Visually, the $\xi^{\kappa_{\rm CMB} \gamma_{\rm t}}(\vartheta)$ estimator seems to show a mild trend for decreasing values of $A$ in lower redshift bins. 
Although such an effect could point towards a  number of interesting phenomena suppressing power for source galaxies at $z\lesssim0.7$
 (e.g. modification to the growth history compared to the fiducial cosmology, additional feedback processes from baryons or massive neutrinos, or redshift-dependent  contamination from IA)
 the significance of this redshift dependence is too low to draw any robust conclusions.
 
 What is  significantly seen from Fig. \ref{fig:A_tomo}   is that the signal is generally  low compared to the fiducial and {\it Planck} cosmologies.
Our measurements of the amplitude $A$ prefer instead the  KiDS-450 cosmology, which also aligns with the  CFHTLenS cosmic shear results \citep{Kilbinger2013}. 
 We further quantify this comparison  in Section \ref{subsec:cosmo_broad}, first presenting results from our set of null tests, and then examining three sources of contamination
 and systematic biases that potentially affect our signal. 
 %
%
%
{\color{black} In this work, we neglect the effect of source-lens coupling \citep{1998A&A...338..375B}, which  could  possibly act as another secondary signal, biasing the signal low.
As it is the case for cosmic shear, this effect should be too small ( $<10\%$) to affect our results significantly, and further investigation will be required to interpret correctly the measurements from future surveys. }

\begin{table}
\caption{$p$-values for the EB test obtained for the 6 tomographic bins. Highlighted in bold is the  $p$-value $\le0.01$. 
The column labelled $C_{10}$ refers to {\small POLSPICE} measurements
in which the data are organized in 10 bins instead of 5. These calculations  assume  $t$-distributed likelihoods \citep[following][]{2016MNRAS.456L.132S}.}
\begin{center}
\begin{tabular}{ccccccc}
\hline
$Z_B$ &  $C^{t{\rm -dist}}$ &   $C_{10}^{t{\rm -dist}}$  & $\xi^{t{\rm -dist}}$   \\
\hline
\hline
\multirow{1}{*}{[0.1, 0.9]}
&0.20 & 0.09 & 0.58 \\
\hline
\hline
\multirow{1}{*}{[0.1, 0.3]}
&\bf 0.01 & 0.09 & 0.52 \\
\hline
\multirow{1}{*}{[0.3, 0.5]}
&0.39 & 0.63 & 0.08 \\
\hline
\multirow{1}{*}{[0.5, 0.7]}
& 0.94 & 0.57 & 0.78\\
 \hline
\multirow{1}{*}{[0.7, 0.9]}
& 0.21 & 0.20 & 0.16\\
\hline
\multirow{1}{*}{$>0.9$}
& 0.53 & 0.54 & 0.68\\
\hline
\end{tabular}
\end{center}
\label{table:p_value_B}
\end{table}%

\subsection{Null tests}
\label{subsec:null}

{\color{black} We have shown in Section \ref{sec:measurement} and in Fig. \ref{fig:cmb_x_shear_KiDS_gamma_t} (red circles) that the parity violation EB test seemed consistent with noise in most
tomographic bins, but occasionally that was not obvious. In this section, we investigate the significance of these measurements.}
Statistically, this is accomplished by measuring the confidence at which we can reject the null hypothesis `{\it parity is not violated}'.
We therefore re-run the full $\chi^2$  statistical analysis\footnote{Due to the absence of parameters in the null hypothesis, the EB case has one additional degree of freedom compared to the EE case.} and measure the $p$-value about the model with $A=0$.
Low $p$-values correspond to high confidence of rejection, i.e. that some residual systematic effect might be causing and apparent parity violation. 
{\color{black} This type of  measurement strongly probes  the tail of the $\chi^2$ distribution, hence assuming a Gaussian likelihood would provide inaccurate estimations of the $p$-values,
even when including the \citet{Hartlap2007a} debiasing $\alpha$ factor. Instead, we follow  \citet{2016MNRAS.456L.132S} and assume a $t$-distribution for the likelihood, 
which better models the tail of the likelihood.
 Table \ref{table:p_value_B} lists all these $p$-values, highlighting in bold one that seems slightly problematic ($p$-value $\le$ 0.01).
 Since this single low $p$-value is only seen in one of the two estimators, we conclude that it must originate from expected noise fluctuations, and 
 not from the data itself.  
This conclusion is additionally supported by the fact that the level of B-modes in the KiDS data (i.e. the BB measurement) 
is consistent with zero on the scales we are probing \citep{KiDS450}.


For the sake of testing the robustness of the EB {\small POLSPICE} measurement, we have additionally investigated the effect of changing the number of bins from 5 to 10.
In the EE case, the recovered values of $A$ and the SNR are similar to those presented in Table \ref{table:stat}, 
from which we conclude that this comes with no gain. 
However, when applied to the EB null test, something interesting happens:
the `problematic' measurement ($p$-value = 0.01 in the $Z_B \in [0.1, 0.3]$ bin, Table \ref{table:p_value_B})
relaxes to 0.09, as  seen in the  column labelled $C_{10}$.
This is another indication that the cause of the low $p$-value  originates from fluctuations in the noise --
which is affected  in the binning process -- without pointing to residual systematic effects in the data.}

We have  verified that our measurement of $A$ is robust against the removal of some scales.
When we exclude the largest or the smallest angular bin in the $\xi^{\kappa_{\rm CMB} \gamma_{\rm t}}$ measurement, 
results change by at most 0.7$\sigma$, generally by less than 0.2$\sigma$. This gives us confidence in the robustness of our measurement.
The same holds when removing the highest $\ell$ bin from the {\small POLSPICE} measurement, but not for the lowest $\ell$ bin,
 which captures the peak of the signal, and therefore contributes significantly to the SNR.
{\color{black} At the same time, this test illustrates that we are currently not sensitive to the effect of massive neutrinos nor to baryonic feedback, 
which affect mainly these non-linear scales.}

 \subsection{Effect of intrinsic alignments}
\label{subsec:IA}

 \begin{figure}
\begin{center}
\begin{tabular}{cc}
\hspace{-5mm}
\includegraphics[width=3.2in]{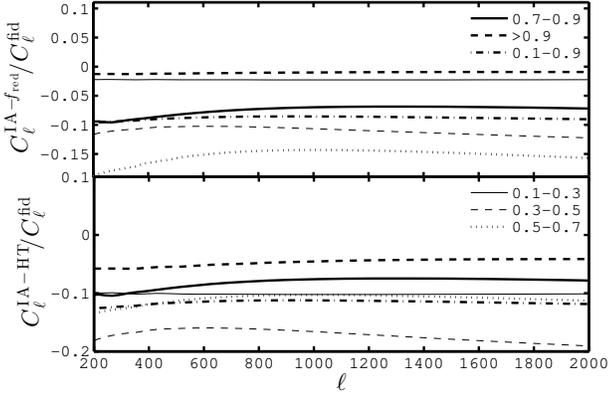}
\end{tabular}
\caption{Strength of the contamination by intrinsic galaxy alignments for different tomographic bins, assuming our fiducial cosmology and
the linear non-linear alignment model.
The difference between lines is caused by changes in $n(z)$ (and in the red fraction  in the $f_{\rm red}$-IA model).} 
\label{fig:frac_IA}
\end{center}
\end{figure}

Intrinsic alignments (IA) are a known secondary effect to the cross-correlation of galaxy lensing and CMB lensing 
that lowers the amplitude of the measured signal \citep{2014MNRAS.443L.119H, TroxelIshak14, 2015MNRAS.453..682C}.
It is therefore important to investigate how much IA could contribute to the observed low values of $A$ reported in Table \ref{table:stat}. 
To estimate the contamination level, we compare two different {\color{black} models}, which we then apply equally to both estimators, $C_{\ell}^{\kappa_{\rm CMB} \kappa_{\rm gal}}$ and $\xi^{\kappa_{\rm CMB} \gamma_{\rm t}}(\vartheta)$. 

First, we follow \citet[][`HT-IA' model hereafter]{2014MNRAS.443L.119H} in using the `linear non-linear alignment' model of \citet{2007NJPh....9..444B}
with the SuperCOSMOS normalization found in \citet{SuperCosmosIA}. {\color{black} We recall that this prescription comes from constraints at $z = 0.1$ that are independent of galaxy type or colour,  and that the effect of IA in this model is to reduce the amplitude of the observed signal, as the  galaxies tend to align radially towards each other.}
The scale-dependence of the alignment contribution is similar to the lensing signal, as seen in \citet{2014MNRAS.443L.119H} and in Fig. \ref{fig:frac_IA}, hence we only quote the
percentage of contamination at $\ell = 1000$ for reference. 
{\color{black} This also allows us to use with confidence the same IA contamination levels for the configuration space estimator, since
rescaling $C_{\ell}^{\kappa_{\rm CMB} \kappa_{\rm gal}}$ by a constant rescales $\xi^{\kappa_{\rm CMB} \gamma_{\rm t}}$ by the same constant (as per equation \ref{eq:limber_cross}). } 
For each of the five redshift bins considered in this paper, starting from the lower redshift, we estimate
a $\{10,17,10, 8, 5\}$\% contamination to the signal, respectively.
For the broader tomographic bin $Z_B \in [0.1, 0.9]$, we estimate a $11\%$ contamination.
In other words, within the HT-IA model, the measured value of $A_{\rm fid}$ in the broad bin (equation \ref{eq:A_fid}) should be corrected to 
\begin{eqnarray}
A_{\rm fid}^{\rm HT} = 0.77\pm0.15
\label{eq:A_fid_HT}
\end{eqnarray} 
The error bars are not modified compared to  the no-IA case since this contamination signal is additive. 
This model is the simplest as it assumes no luminosity or redshift dependence of the alignment normalization, and adopts the same alignment prescription for all galaxies regardless of morphological type/colour.

Second, we estimate the contamination from the alignment model of \citet[][`$f_{\rm red}$-IA' model hereafter]{2015MNRAS.453..682C} that allows for differential contributions based on galaxy colour/morphology.
We assume that blue galaxies do not contribute at all, consistent with observations \citep{Heymans2013a, BlueIA}, even though this null measurement remains poorly constrained.
We estimate the red fraction directly from the data in each redshift bin, using the BPZ template information {\tt T\_B}.
Motivated by \citet{2015MNRAS.449.1505S}, we identify red galaxies as objects with {\tt T\_B} $<1.5$. 
For the five tomographic bins,  we obtain  fractions of red galaxies $f_{\rm red} = \{0.04, 0.12, 0.27, 0.18, 0.04\}$; we estimate $f_{\rm red} = 0.15$ for the broad bin.
We then use the alignment amplitude for the red galaxies from \citet{Heymans2013a} to obtain an estimate of alignment contamination given our red fractions.
These results are presented in the upper panel of Fig. \ref{fig:frac_IA}. 
With this method, we estimate a $\{2, 11, 14, 7, 1\}$\% contamination from intrinsic alignments in the tomographic bins, 
and a 9\% contamination in the $Z_B \in [0.1, 0.9]$ bin\footnote{
We measured the field-to-field variance in $f_{\rm red}$ and observed that it hardly varies except in the highest redshift bin, where
the scatter could turn the 1\% IA contamination into a 0.0 -- 2.5\% contamination. 
This remains small and should have a negligible  impact.}.
Then we can estimate  
\begin{eqnarray}
A_{\rm fid}^{f_{\rm red}} = 0.75\pm0.15.
\label{eq:A_fid_fred}
\end{eqnarray} 
{\color{black} One caveat with this model is that the $K$-correction and evolutionary corrections are uncertain at high redshift, 
which could result in biased estimates of the red fraction \citep[see discussion in][]{2015MNRAS.453..682C}.
This has an impact on the exact level of intrinsic alignment contamination by red galaxies, but we neglect this effect in this work.}

Both methods are broadly consistent even though they differ in details, especially in the lowest redshift bin.
 For instance, the second method captures the redshift differences observed in the data and takes into account the split in contributions arising from different galaxy types,
which introduces a slightly different redshift dependence of the IA signal.
The overall trends between the HT and the $f_{\rm red}$-IA models are similar though, but that is not the case for all IA models 
\citep[see, for example, the tidal torque theory from][in which the sign of the IA effect on the signal is the opposite]{2015MNRAS.452.3369C}.
There remains a large uncertainty in the modelling of the IA contamination, and we do not know which model, if any,
should enter in our cosmological interpretation.

According to these estimations, both the HT-IA and $f_{\rm red}$-IA models help to bring $A$ closer to unity.
From the contamination levels listed above, at most 17\% of the observed cross-correlation signal can be canceled by IA contamination in our tomographic bins. 
After correcting for this effect in each tomographic bin, most points agree with $A_{\rm fid}=1$ within $1\sigma$.
This is shown with the color symbols in Fig. \ref{fig:A_tomo}.

Finally, we note that the uncertainty on the level of IA contamination quoted in the section is high, 
especially because of the unknown signal from the blue galaxies.
For instance, at the $1\sigma$ level and assuming the linear non-linear alignment model, the IA contamination from blue galaxies could range from 
$-\{10, 15, 8, 6, 4 \}$\% to +$\{6, 9, 5, 4, 3\}$\% in each tomographic bin, and from $-10\%$ to +6\% in the $Z_B \in [0.1, 0.9]$ bin.

%

\subsection{Effect of $n(z)$ errors}
\label{subsec:nz}

\begin{figure}
\begin{center}
\begin{tabular}{cc}
\includegraphics[width=3.2in]{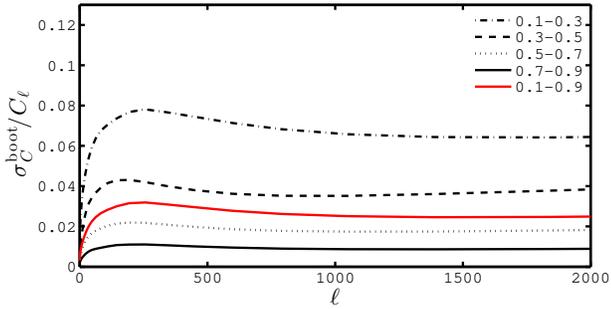}\\
\end{tabular}
\caption{Fractional effect on the $C_{\ell}^{\kappa_{\rm CMB} \kappa_{\rm gal}}$ signal when varying the $n(z)$ between 100 bootstrap resamplings, for the four tomographic bins with $Z_B<0.9$. 
Shown is the $1\sigma$ scatter divided by the signal. } 
\label{fig:frac_nz}
\end{center}
\end{figure}

We investigate here the impact on our measurement of $A$ from the uncertainty on the source redshift distribution.
This is estimated from 100 bootstrap resamplings of the source catalogue, as detailed in \citet[][the DIR method described therein]{KiDS450}. 
These samples consist of internal fluctuations in the $n(z)$, which we turn into fluctuations in the signal with equations \ref{eq:limber_cross} - \ref{eq:g-chi-def}. 
We present in the top panel of Fig. \ref{fig:frac_nz} the fractional error on the signal, i.e. $\sigma^{\rm boot}_{C_{\ell}}/C_{\ell}$. 
According to this error estimate, the uncertainty on $n(z)$ is up to 8\% of the signal in the  first redshift bin, then 4, 2 and 1\% for the others, {\color{black} and about 3\% 
for the $Z_B \in$ [0.1, 0.9] tomographic bin.}

Note that this quantity is a measure of how the DIR $n(z)$ varies -- and how it impacts the signal -- across subsamples of the re-weighted spectroscopically matched catalogue.
This catalogue is by itself subsampling the full KiDS sources, and hence subject to sampling variance. 
It is therefore likely that the error quoted above slightly  under-estimates the true error on the signal due to the $n(z)$,
{\color{black} as discussed in Section C3.1 of \citet{KiDS450}}.
This is sub-dominant compared to our statistical uncertainty and is therefore not expected to affect our results.

For comparison purposes, we also investigated estimates of the redshift distribution determined using the cross-correlation between spectroscopic and photometric samples 
\citep[known as the CC method in][]{KiDS450,TheWiZZ}.  This scheme has a high level of noise compared to the fiducial DIR method and we find that the error on the recovered $C_\ell$ in our analysis increases from $\sim 5\%$ in the DIR case to $\sim 30\%$ in the CC case.    From this we can draw the same conclusion as the KiDS-450 cosmic shear analysis, that determining the redshift distribution using the cross-correlation CC method will remove any  discrepancy with a {\it Planck} cosmology through the inflation of the error bars.  We believe however that the error on the CC estimate is not representative of our actual knowledge of the $n(z)$ in the KiDS data, and refer instead to the redshift distribution defined using DIR method in the rest of this paper.

Precision on the KiDS source redshift distribution will soon increase thanks to the ongoing processing of near-IR VIKING data (Hildebrandt et al. in prep.), 
which primarily impact the high redshift tail so crucial to our measurement. Finally note that in the DIR method we are using a calibrated $n(z)$, estimated from weighted spectroscopic data, 
hence we do not have to worry nearly as much about catastrophic photo-$z$ outliers. This was not the case for the analysis presented in HD16, which showed 
that for $n(z)$ estimated directly from photometric data (for e.g. CFHTLenS and RCSLenS), these can easily dominate the error budget, with systematic effects on the signal of the order of 15\%.
If our measurement contains more high-redshift objects than our $n(z)$ suggests, our predictions are too low; correcting for this would lower $A$.

\subsection{Baryon feedback, massive neutrinos and non-linear modelling}
\label{subsec:baryon}

 \begin{figure}
\begin{center}
\begin{tabular}{cc}
\hspace{-5mm}
\includegraphics[width=3.2in]{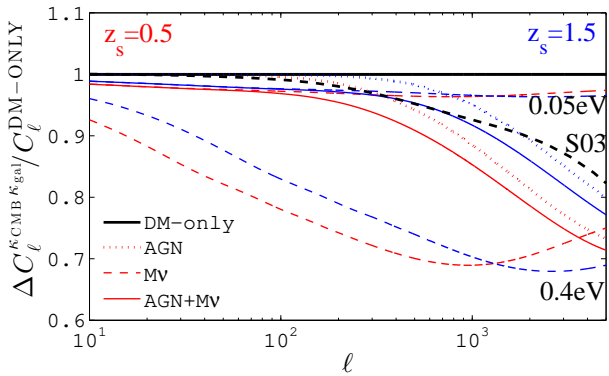}
\end{tabular}
\caption{Fractional effect of the AGN baryon feedback and massive neutrinos on the cross-spectra, for different combinations of source planes.
The red solid line shows the combined effect on the cross-spectrum for sources placed on a single plane at $z_s = 0.5$.
The effect of 0.05eV massive neutrinos and AGN feedback are shown separately by the upper dashed and the dotted line (also in red). 
The lower dashed redline shows the impact of 0.4eV neutrinos.
Blue lines show the same quantities, but  for sources placed at $z_s = 1.5$.
{\color{black}The dashed black line shows the ratio between the predictions from \citet{Smith2003} and that of \citet{Takahashi2012}.}} 
\label{fig:frac_baryons}
\end{center}
\end{figure}

As shown in HD16, baryonic feedback and massive neutrinos can cause an important decrease of the cross-correlation signal, 
which would translate into lower values of $A$ when compared to a fiducial dark matter only cosmology.
To investigate how this could affect our cosmological results, we modify the $P(k,z)$ term in equation \ref{eq:limber_cross} to include  `massive  neutrino bias' and `baryon feedback bias' as detailed in \citet{HWVH2015}. 
The baryon bias was extracted from the OWL simulations, assuming the  AGN model \citep{vanDaalen2011},
while the neutrino bias was extracted from the recalibrated {\small HALOFIT} code \citep{Takahashi2012} with {\color{black} total} neutrino masses ${\color{black} M_{\nu}}$ = 0.05, 0.2, 0.4 and 0.6eV. 
Our results are presented in Fig. \ref{fig:frac_baryons} for two simplified cases, 
in which the source galaxy populations are placed on single planes at $z_s = 0.5$ (in red) and at $z_s = 1.5$ (in blue). 
The figure focuses on the 0.05eV scenario, showing the suppression of power caused by massive neutrinos (3.5\% effect on $A$ for both $z_s$ planes, averaged over the $\ell$-modes that we measured),
by baryonic feedback (5.0\% for  $z_s = 1.5$ and 10.6\% for $z_s = 0.5$), and by the combination of both (8.2\% and 13.7\% for $z_s = 1.5$ and $0.5$ respectively).
The reason why the effect of baryons is larger on the lower redshift source plane is simply a projection effect: the same physical scales subtend different angles on the sky,
which contribute differently to our measurement restricted to the $\ell \in [20 -2000]$ range.
We also show the effect of 0.4eV neutrinos (28\% and 30\% in the two $z_s$ slices), which demonstrates a scaling of 7\% per 0.1eV.   
We note that \citet{MeadFit} proposes an alternative method to account simultaneously for baryons and neutrinos based on the halo model, 
which might prove useful in future work.

These two effects contribute at some level to the measurement of $A$, {\color{black} but it is too early to put constraints on them based on our measurement. Firstly,} 
  the cosmology is not guaranteed to be that of {\it Planck}, secondly the exact feedback mechanism that is at play in the Universe remains largely unknown,
and thirdly other effects (e.g. IA contamination or error in the $n(z)$) could explain why our measured $A_{\rm fid}$ is low.
However, if the fiducial cosmology is correct and if the intrinsic alignments are well described by the HT model described in Section \ref{subsec:IA},
then $A_{\rm fid}$ would be brought to unity with $M_{\nu} = 0.33{\color{black}\ \pm\ 0.22}$eV in absence of baryonic feedback, and $M_{\nu} = 0.19{\color{black}\ \pm\ 0.22}$eV within the AGN model.

{\color{black} We have verified that the uncertainty in the non-linear modelling does not affect our measurement of $A$ significantly. 
This is mainly because the angles and redshifts probed by our measurement correspond to scales that are mostly in 
the linear and mildly non-linear regime. Replacing the non-linear power spectrum from the \citet{Takahashi2012} model
with that of \citet{Smith2003}, a radical change in the non-linear predictions at small scales {\color{black} shown in Fig. \ref{fig:frac_baryons}} (black dashed line), affects our measurement of $A$ by 1-2\% only.
This is well within the statistical uncertainty and can be safely neglected. 
 Fig. \ref{fig:frac_baryons} shows that there is a clear degeneracy between differences in the two models, and the effect of baryonic feedback. 
However, the \citet{Smith2003} predictions are known to suffer from a significant loss of power at small scales, 
visible in Fig. \ref{fig:frac_baryons}, and the state-of-the-art precision on the non-linear power spectrum, from e.g. the Cosmic Emulator \citep{Heitmann2013}  deviates from the \citet{Takahashi2012} model by less than 5\%  \citep{MeadFit}. 
This alleviates the degeneracy between modelling and baryonic feedback effects and further supports  our (model-dependent) neutrino mass contraints presented above. }

\subsection{Cosmology from broad $n(z)$}
\label{subsec:cosmo_broad}

\begin{figure}
\begin{center}
\begin{tabular}{cc}
\hspace{-5mm}
\includegraphics[width=3.5in]{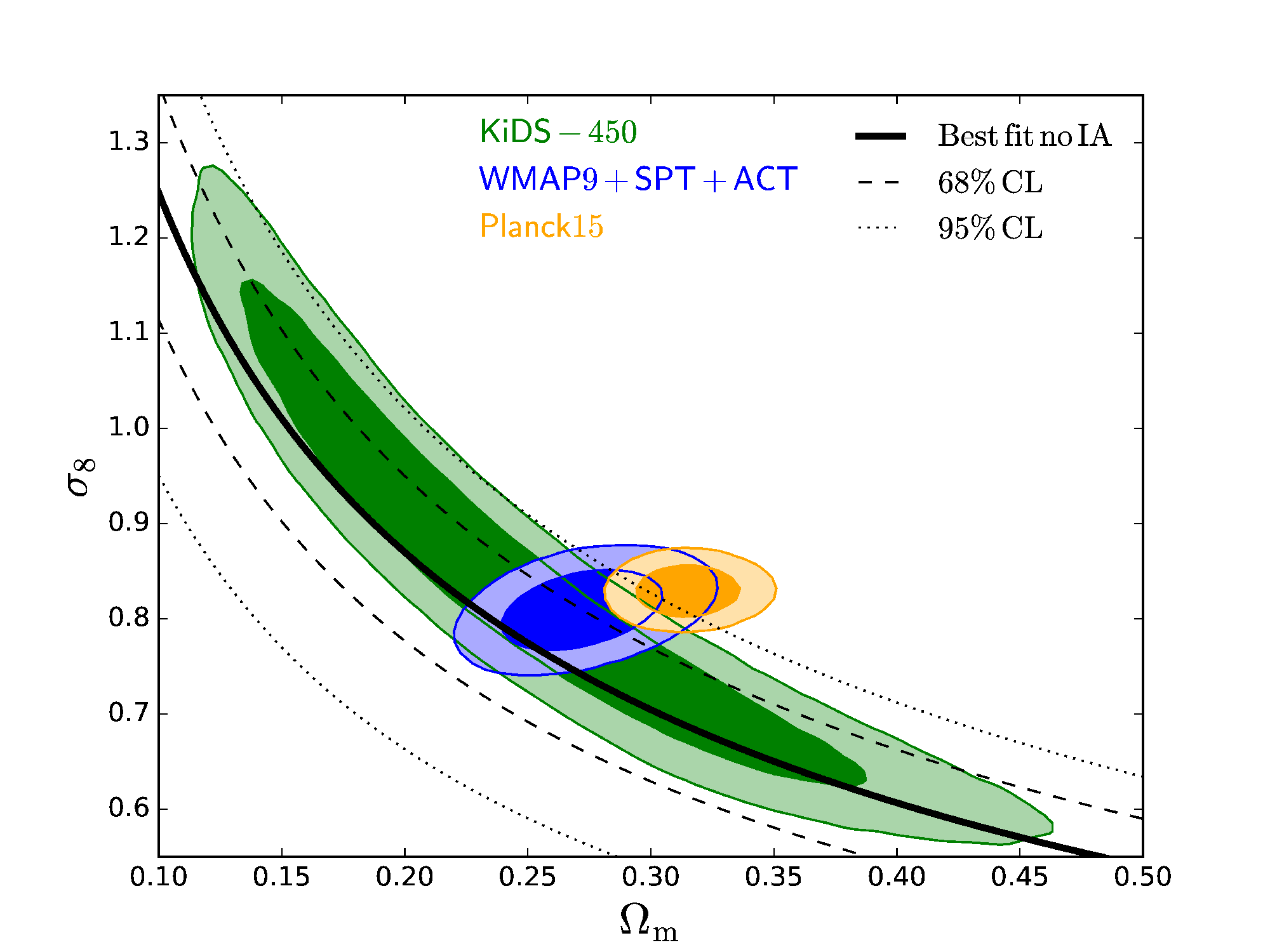}
\end{tabular}
\caption{Constraints on $\sigma_8$ and $\Omega_{\rm m}$ as estimated from  the cross-correlation measurement, ignoring potential contamination by intrinsic galaxy alignements 
(shown in black).  The solid line shows the best fit, while the dashed and dotted lines indicate the  68\% and 95\% confidence level (CL) regions, respectively. 
The cross-correlation results can be compared to KiDS-450 (green, where IA are accounted for) {\it Planck} (orange) and WMAP9+SPT+ACT (blue).} 
\label{fig:Cosmo}
\end{center}
\end{figure}

\begin{figure}
\begin{center}
\begin{tabular}{cc}
\hspace{-5mm}
\includegraphics[width=3.5in]{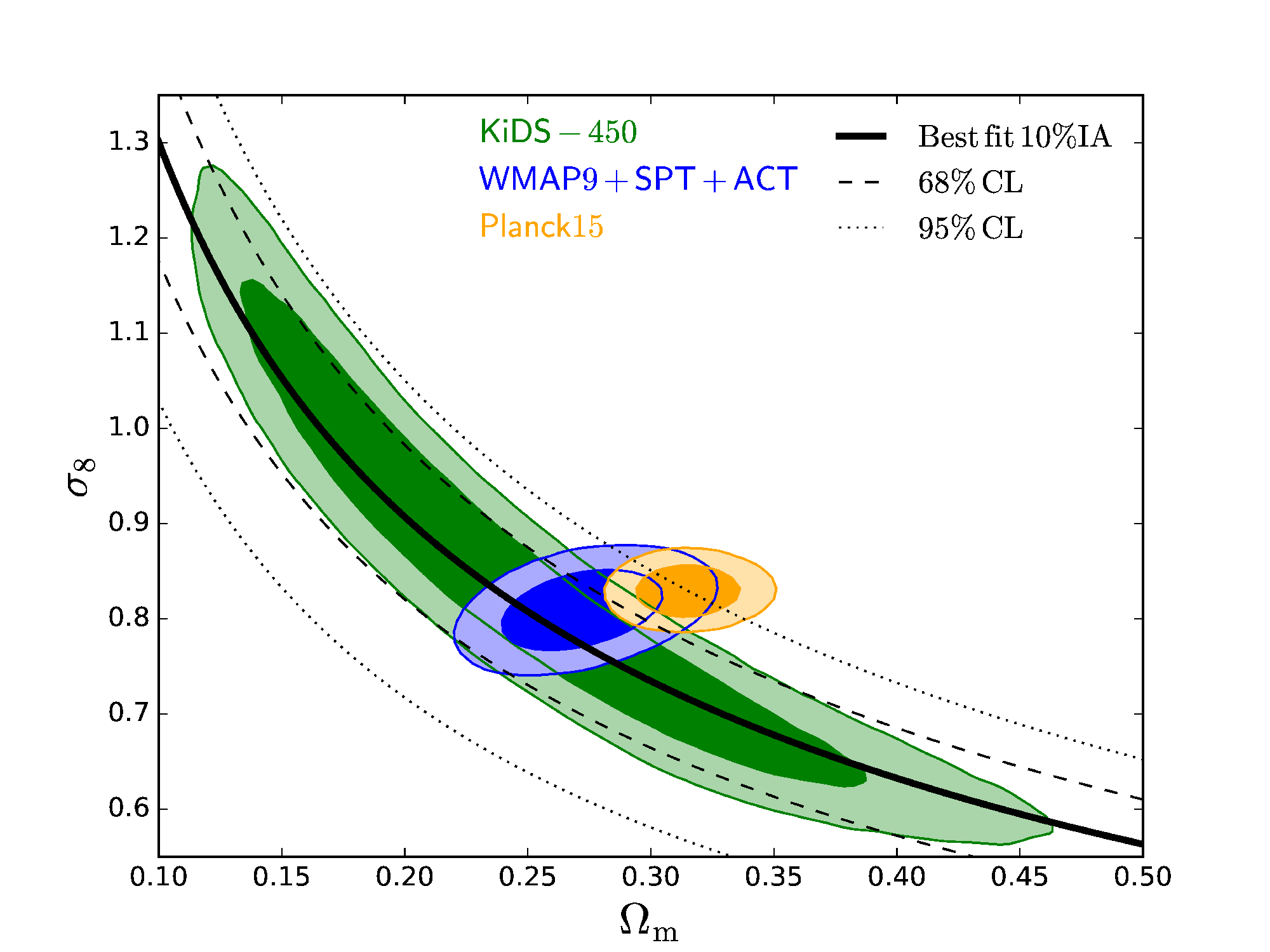}
\end{tabular}
\caption{Same as Fig. \ref{fig:Cosmo}, but here assuming 10\% contamination from IA in the cross-correlation measurement (equation \ref{eq:A_cosmo_HT}), consistent with both the `HT-IA' and the `$f_{\rm red}$-IA'  models. } 
\label{fig:Cosmo_IA}
\end{center}
\end{figure}

In this section, we investigate  how our cross-correlation measurement can constrain cosmology, and specifically compute confidence regions
in the $[\sigma_8- \Omega_{\rm m}]$ plane. For this calculation we assume massless neutrinos, no baryonic feedback, we ignore the error on $n(z)$, but examine
our results for the three IA models described in Section \ref{subsec:IA}.

It was shown in  \citet{2015PhRvD..92f3517L} that the amplitude of the cross-correlation signal scales approximately with 
$[\sigma_8^{2} \Omega_{\rm m}^{-0.5}]$ at large, linear scales ($\ell<$  few hundred), and as $[\sigma_8^{3} \Omega_{\rm m}^{1.3}]$ at small scales ($\ell>$ 1000).
Most of our constraints come from small scales, but our measurement includes some large modes down to $\ell\sim200$. 
For this reason,  we strike a compromise: we keep the $\Omega_{\rm m}^{1.3}$ dependence, as suggested by   \citet{2015PhRvD..92f3517L},
but use a $\sigma_8^{2.5}$ dependence, to capture the gradual transition between both.  
Future measurements will require MCMC algorithms to be run to better capture these dependencies, 
but this does not seem to be necessary in this case given the relatively large uncertainty on $A$.

As discussed before, we use the $\xi^{\kappa_{\rm CMB} \gamma_{\rm t}}$ results in the broad $Z_B \in [0.1, 0.9]$ tomographic bin
because it has the highest signal to noise, however our results would not change significantly  if we used the $C_{\ell}^{\kappa_{\rm CMB} \kappa_{\rm gal}}$
measurement instead. We could also have used the tomographic results, i.e. the $A(z)$ in the four bins. 
However these measurements are all correlated, probing common low-redshift lenses. 
This would require us to calculate and include cross-correlation coefficients between the different tomographic bins
when solving for the best-fit cosmology. These could be evaluated from mock data, but this is not required when working with a single data point for $A$.
Combining this scaling relation with equations \ref{eq:A_fid} - \ref{eq:A_fid_HT}, we get 
%
\begin{eqnarray}
A = A_{\rm fid} \left(\frac{\sigma_8}{0.831}\right)^{2.5} \left(\frac{\Omega_{\rm m}}{0.2905}\right)^{1.3} = 0.69\pm0.15
\label{eq:A_cosmo}
\end{eqnarray}
and 
\begin{eqnarray}
A = A_{\rm fid}^{\rm HT} \left(\frac{\sigma_8}{0.831}\right)^{2.5} \left(\frac{\Omega_{\rm m}}{0.2905}\right)^{1.3} = 0.77\pm0.15, 
\label{eq:A_cosmo_HT}
\end{eqnarray}
which we use to propagate the error on $A$ into confidence regions in the $[\sigma_8 - \Omega_{\rm m}]$ plane. 
We show in Fig. \ref{fig:Cosmo} and \ref{fig:Cosmo_IA} how these constraints compare to the results from KiDS-450 cosmic shear {\color{black} (with IA)}, {\it  Planck} and other CMB experiments\footnote{\color{black} The MCMC chains entering these contour plots can be found on the KiDS-450 website: {kids.strw.leidenuniv.nl/cosmicshear2016.php}. Note also that the WMAP9+SPT+ACT cosmology presented in Fig.   \ref{fig:Cosmo} and \ref{fig:Cosmo_IA} differ from the fiducial WMAP9+BAO+SN cosmology.}. 
 Our cross-correlation measurement has a larger overlap with the KiDS-450 constraints but is still consistent with the {\it Planck} cosmology in the sense that their $\lesssim 95\%$ confidence regions overlap. Including IA reduces the offset from {\it Planck}.

Given that our signal has different dependences on cosmological (e.g. $\Omega_{\rm m}$, $\sigma_8$) and nuisance (e.g. $m$, $n(z)$) parameters,
we can see how this can provide new insights in resolving tensions between the cosmic shear and CMB measurements.
For example, whereas the KiDS-450 and CFHTLenS cosmic shear results scale as $[m^2 n^{2}(z)]$, 
our KiDS-450 $\times$ {\it Planck} lensing measurement scales as $[m\,n(z)]$.
This difference could therefore allow us to break the degeneracy in a joint probe analysis. 
Also note that in general, we should not exclude possibility that there could be residual systematics left over in a CMB temperature and polarization analysis -- driving the cosmology to higher $[\sigma_8, \Omega_{\rm m}]$ values --
that do not make their way to the CMB lensing map or into the joint probes measurement, in analogy with the cosmic shear $c$-term. This is certainly the case for the additive shear bias (the $c$-correction) described in \citet{KiDSLenS}. 
Having this new kind of handle can help to identify the cause of disagreements between different probes, and will be central to the cosmological analyses of future surveys.
We explore further how cross-correlation analyses can be turned into a calibration tool in the  next section. 

 \subsection{Application: {photo-$z$} and $m$-calibration}
\label{subsec:m}

The {\color{black} CMB lensing -- galaxy lensing cross-correlation signal} has been identified as a promising alternative to calibrate the cosmic shear data without relying completely on image simulations
\citep{2013arXiv1311.2338D, 2016PhRvD..93j3508L, 2016arXiv160701761S}. This statement relies on the fact that $A$ absorbs all phenomena that affect the amplitude of the measurement, 
i.e. cosmology, intrinsic alignment, $n(z)$, shear calibration,  and that we can marginalize over some of these in order to solve for others.

\begin{figure}
\begin{center}
\begin{tabular}{cc}
\hspace{-5mm}
\includegraphics[width=3.2in]{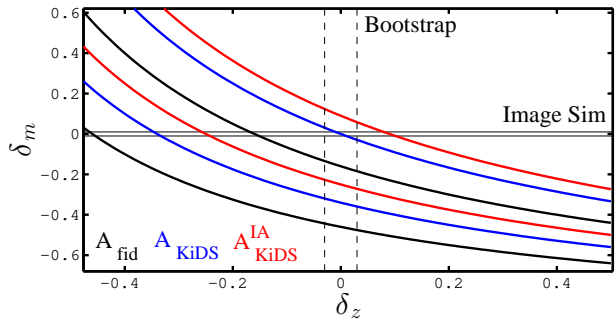}
\end{tabular}
\caption{$1\sigma$ contour regions on the shear calibration correction $\delta_m$ and the redshift distribution correction $\delta_z$ derived from the  $Z_B \in [0.1, 0.9]$ measurement of $A$ in three different cases.
Results from $A_{\rm fid}$ (no-IA) are shown in solid black, results from $A_{\rm KiDS}$ (no IA) are shown in solid blue, 
and results  from   $A_{\rm KiDS}$ with 10\% IA are shown in solid red.
The pair of solid horizontal lines show the region of $\delta_m$ values allowed from image simulations,
while the pair of dashed vertical lines show the region of $\delta_z$ values allowed from bootstrap resampling the $n(z)$. } 
\label{fig:deltam_deltaz_low_z}
\end{center}
\end{figure}

Most of the attention so far has been directed towards the multiplicative term in the cosmic shear  calibration -- the $m^{j}$ factor in equation \ref{eq:K} -- which 
has an important impact on the cosmological interpretation. In the case of the KiDS-450 data, the shear calibration is known at the percent level from image simulations
for objects with $Z_B<0.9$ (see Fig. \ref{fig:deltam_deltaz_low_z}), but the precision on $m^{j}$ quickly degrades at higher redshift \citep{2016arXiv160605337F}.
Similar conclusions can be drawn from the photometric redshift estimation, which becomes unreliable at high redshift when only using optical bands \citep{KiDS450}. 
We see in our cross-correlation measurement a unique opportunity to place a joint-constraint on these two quantities in the highest  redshift bin, 
informed by our measurement at lower redshift. {\color{black} We ignore the contribution from IA, knowing that this will need to be included for upcoming surveys\footnote{
{\color{black} One might well object that the uncertainty in IA modelling and its evolution is already larger than the uncertainty in shear calibration, 
and hence that our strategy is flawed to start with. Instead, we should be placing simultaneous constraints on the photo-$z$, $m$-calibration and IA. 
A full MCMC will certainly be required in the future to disentangle these effects, exploiting their different shape dependence to break the degeneracy between these parameters. 
As an illustration of this strategy however, we present a simple case and assume no IA contamination in this rest of this section.}}. }

Our approach is to fix the scaling term $A$ to the value preferred by the  $Z_B \in [0.1, 0.9]$ data, which we label $A_{{\rm low}z}$ here for clarity, and to 
jointly fit for the mean shear bias  and mean redshift distribution in the $Z_B>0.9$ bin.
Forcing $A$ to this value in the high redshift bin provides constraints on $\langle m_{{\rm high}z}\rangle$ and $\langle n_{{\rm high}z}(z)\rangle$, 
which we extract by varying these quantities in the predictions.

The correction to the shear bias is trivial to implement as it scales linearly with $A$,
so we simply write $A_{{\rm high}z} = A_{{\rm low}z}(1 + \delta_{m}) = 0.95 \pm 0.22$ (from Table \ref{table:stat}) and solve for $\delta_{m}$.
If this was the only correction, we could write $\delta_{m} = A_{{\rm high}z}/A_{{\rm low}z} - 1 = 0.38\pm0.44$, 
which is consistent with zero but not well constrained.

Corrections to the photometric distribution can be slightly more complicated as the full redshift distribution that enters our calculation is not simple, as seen in Fig. \ref{fig:nz}.
There are a number ways with which we could alter the $n(z)$ and propagate the effect onto the signal, e.g. by modifying the overall shape, the mean or the tail of the distribution. 
We opted for arguably the simplest prescription, which consists in shifting the $n(z)$ along the $z$ direction by applying the mapping $z\rightarrow z + \delta_z$ 
(thereby shifting  $\langle n_{{\rm high}z}(z) \rangle$ by the same amount). We propagate this new $n(z)$ through equation \ref{eq:limber_cross} and solve for values of $\delta_z$ that satisfy constraints on $A$. 
In this process, we allow $\delta_z$ to vary by up to 0.5, which is rather extreme. 
%

\begin{figure}
\begin{center}
\begin{tabular}{cc}
\hspace{-5mm}
\includegraphics[width=3.2in]{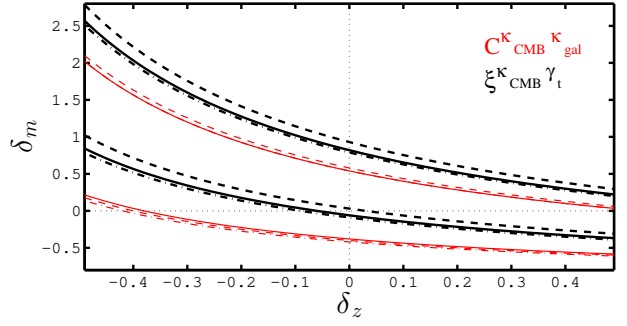}
\end{tabular}
\caption{$1\sigma$ contour regions on the shear calibration correction $\delta_m$ and the redshift distribution correction $\delta_z$  in the bin $Z_B >0.9$,  from the cross-correlation measurements.
Black and red correspond to  constraints from  $\xi^{\kappa_{\rm CMB} \gamma_{\rm t}}$ and $C_{\ell}^{\kappa_{\rm CMB} \kappa_{\rm gal}}$ respectively.
The multiple lines present the results in three different cosmologies (fiducial is solid, KiDS-450 is dot-dashed, {\it Planck} is dashed), which are shown here to have a small impact on the constraints.
Other independent measurements and improved image simulations could tighten the region of allowed values. The upper red solid and dot-dashed lines perfectly overlap.
} 
\label{fig:deltam_deltaz}
\end{center}
\end{figure}

Following the simple reasoning described for the  shear calibration, we can see that if $m$ was trusted at the percent level in this high redshift bin, 
constraints on the redshift distribution could be simply derived by computing $A_{{\rm high}z}/A_{{\rm low}z} = C_{\delta_z} / C_{\rm fid} = (1+\delta_z)$. 
We therefore obtain the exact same constraints as for $\delta_m$, namely:  $\delta_z=0.38\pm0.44$.
We place constraints on the [$\delta_m - \delta_z$] plane by requiring $(1+\delta_m)(1+\delta_z) = {A_{{\rm high}z}}/{A_{{\rm low}z}}$,
and present the $1\sigma$ constraints in Fig. \ref{fig:deltam_deltaz}. The data are still consistent with $\delta_m = \delta_z = 0$,
but these two biases are not currently well constrained. 

{\color{black} We also show in Fig. \ref{fig:deltam_deltaz}  the results from the $C_{\ell}^{\kappa_{\rm CMB} \kappa_{\rm gal}}$ estimator (in red dashed),
but these have a lower SNR hence are not included in the analysis. At first sight, the difference observed  between the results from the two estimators could seem worrisome.
Given that these constraints on [$\delta_m - \delta_z$] are obtained from the same data, and that the only difference is the analysis method, 
it is justified to question whether we could use this measurement for precise self-calibration if two methods on the same data give such different values for $\delta_m$ and $\delta_z$.
We recall that differences are expected since both techniques are probing different scales, however the calibration technique presented here is sensitive to these differences. 
The calibration is only weakly sensitive to the fiducial cosmology adopted, as shown in Fig. \ref{fig:deltam_deltaz}  with the different line styles.

A significant  improvement will come from the future data sets (advanced-ACT, SPT-3G, LSST, Euclid), in which the noise will be much lower, allowing for more accurate measurements of 
$\xi^{\kappa_{\rm CMB} \gamma_{\rm t}}$ and $C_{\ell}^{\kappa_{\rm CMB} \kappa_{\rm gal}}$ to start with. 
In addition, including other measurements in this self-calibration approach will greatly enhance the achievable precision.}
%
{\color{black} For example, one could measure the galaxy-galaxy lensing signal
 from the same KiDS-450 source galaxies, using  i.e. the GAMA galaxies as lenses \citep{2016MNRAS.459.3251V},
selecting the sources in the same tomographic bin (i.e. $Z_B \in  [0.1, 0.9]$ and $Z_B>0.9$).
Fixing the cosmology from the low redshift bin, one could then similarly constrain [$\delta_m - \delta_z$] in the high redshift bin.
The idea here is that the trend can be made {\it opposite} to that seen in Fig. \ref{fig:deltam_deltaz}: an increase in $\delta_z$ pushes the sources away from the lenses, which, 
depending on the geometry, could reduce the signal. To compensate for this, the $m$-calibration would need to increase as well. 
In such a setup, the preferred region in parameter space would inevitably intersect with ours, and exploiting this complementarity might lead to  competitive constraints.
Further investigation on this combined measurement will be explored in upcoming work.}

We are aware that our bi-linear modelling of the $m$ and $n(z)$ calibration  is an over-simplification of our knowledge (and uncertainty) 
about these quantities in the highest redshift bin, and one could envision {\color{black} improving this strategy} in the future.
 For instance, the high-redshift objects are often the hardest to measure spectroscopically, hence there are higher chances that the DIR method fails at higher redshifts. 
To capture this effect, instead of shifting the $n(z)$, one could modify only the high-redshift tail, moving 1\%, 5\% or 10\% of our source galaxies 
from (very) low redshifts to $z>$1, propagating the effect on the signal, and use our measurement of $A$ to constrain the fraction of such `missing' high-redshift galaxies.   
However, given the size of our error bars, it is not clear that we would learn more from this approach at the moment. 

{\color{black} This situation will improve significantly with future CMB and galaxy surveys. 
According to \citet{2016arXiv160701761S},  the lensing data provided by a Stage-4 CMB experiment, combined with 
10 tomographic bins for LSST, will enable a $m$-calibration that is accurate to better than 0.5\%. 
This is marginalising over a number of nuisance parameters that unfortunately does not include catastrophic photometric redshift outliers, 
so the actual accuracy will likely degrade compared to this impressive benchmark. 
Nevertheless, this is an avenue that is certainly worth exploiting with the upcoming data.}

{\color{black} The $Z_B<0.9$ redshift data in the KiDS survey has been calibrated on image simulations whose precision on $\delta_m$
largely surpasses that of the cross-correlation technique presented in this section.
Fig. \ref{fig:deltam_deltaz_low_z} shows the $1\sigma$ constraints in the $\left[\delta_m - \delta_z \right]$ plane in the $Z_B \in [0.1, 0.9]$ bin assuming the fiducial cosmology without IA (black), 
the KiDS-450 cosmology without IA (blue), and the KiDS-450 with 10\% IA (red), consistent with both the HT-IA and the $f_{\rm red}$-IA models.
For comparison, the 1\% precision on $\delta_m$ obtained from image simulations  and the 3\% precision on $\delta_z$ obtained from bootstrap resampling the $n(z)$, described in Section \ref{subsec:nz},
are shown as the pairs of horizontal and vertical lines, respectively. For these redshifts at least, the measurement provides interesting constraints on the cosmology, IA and $\delta_z$, but not on the  $m$-calibration. }

\section{Conclusions}
\label{sec:conclusion}

We perform the first tomographic lensing-lensing cross-correlation by combining the {\it Planck} 2015 lensing map with the KiDS-450 shear data.
Our measurement is based on two independent estimators, the {\small POLSPICE} measurement of $C_{\ell}^{\kappa_{\rm CMB} \kappa_{\rm gal}}$, 
and the configuration-space measurement of $\xi^{\kappa_{\rm CMB} \gamma_{\rm t}}(\vartheta)$. 
The two techniques agree within $1\sigma$ in all tomographic bins, although the former exhibits a lower signal to noise ratio. 

{\color{black} We compare our tomographic results against a two-dimensional lensing analysis of a single broad redshift bin ($Z_B \in [0.1, 0.9]$),
and fit the measured amplitude of the signal with a single multiplicative parameter $A$ that scales the predictions.
We obtain $A_{\rm fid} = 0.69\pm0.15$ in our fiducial cosmology,
and show that the constraints on the $[\sigma_8 - \Omega_{\rm m}]$ plane are consistent with the  flat $\Lambda$CDM {\it Planck} cosmology 
at the 95\% level, with $A_{\it Planck} = 0.68\pm0.15$, and with all previous results  \citep[][K16 \& HD16]{2015PhRvD..91f2001H, 2015PhRvD..92f3517L, 2017MNRAS.464.2120S}. 
The KiDS-450 cosmology is preferred however, in which we obtain $A_{\rm KiDS} = 0.86\pm0.19$.}


Photometric redshifts have been examined carefully and are unlikely to be affecting these results significantly ($<8\%$ effect on the signal), unless the spectroscopic sample that is used to estimate the $n(z)$ suffers from significant sampling variance. Multiplicative shear calibration is also highly unlikely to be affecting $A$ since it is known to be accurate at the percent level over the redshift range that enters our cosmological measurement.  However, including different models of intrinsic alignment, massive neutrinos and baryon feedback in the predictions all affect the signal by tens of percent, pushing the recovered $A$ {\color{black} to higher values}.


Fixing the cosmology to that favoured by our low-redshift measurements ($Z_B < 0.9$), we calibrate the high-redshift ($Z_B>0.9$) photometric $n(z)$ 
and the multiplicative shear calibration,  which are {\color{black} not robustly constrained}. 
We find that the high redshift data is consistent with no residual systematics, but that these are still allowed and only weakly constrained. 
Improved results on this high redshift calibration will come in the future from larger data sets, from improved image simulations and from the combination with other independent measurements.

Tomographic measurements such as that presented in this paper are insensitive to galaxy bias, 
and hence opening the possibility to obtain cosmological constraints from measurements of the growth factor.
Upcoming and future lensing surveys will have excellent opportunities for combining probes and improving their cosmological 
analyses.

\section*{Acknowledgements}


Based on data products from observations made with ESO Telescopes at the La Silla Paranal Observatory under programme IDs 177.A-3016, 177.A-3017 and 177.A-3018, and on data products produced by Target/OmegaCEN, INAF-OACN, INAF-OAPD and the KiDS production team, on behalf of the KiDS consortium. OmegaCEN and the KiDS production team acknowledge support by NOVA and NWO-M grants. Members of INAF-OAPD and INAF-OACN also acknowledge the support from the Department of Physics \& Astronomy of the University of Padova, and of the Department of Physics of Univ. Federico II (Naples).

JHD acknowledges support from the European Commission under a Marie-Sk{\l}odowska-Curie European Fellowship (EU project 656869).
TT, LvW and AH are supported by the NSERC of Canada, TT is additionally funded by CIFAR.
HHi is supported by an Emmy Noether grant (No. Hi 1495/2-1) of the Deutsche Forschungsgemeinschaft.
MV,  CH and MA acknowledge support from the European Research Council under FP7 grant number 279396 (MV) and 647112 (CH, MA). 
MV is also supported by the Netherlands Organisation for Scientific Research (NWO) through grants 614.001.103.
Parts of this research were conducted by the Australian Research Council Centre of Excellence for All-sky Astrophysics (CAASTRO), through project number CE110001020.
JM has received funding from the People Programme (Marie Curie Actions) of the European Union's Seventh Framework Programme (FP7/2007-2013) 
under REA grant agreement number 627288. ENC is supported by a Beecroft fellowship. PS is supported by the Deutsche
Forschungsgemeinschaft by the TR33 `The dark Universe'. LM is supported by STFC grant ST/N000919/1, and FK by a de Sitter Fellowship of the Netherlands Organization for Scientific Research (NWO) under grant number 022.003.013. KK acknowledges support by the Alexander von Humboldt Foundation.


{\footnotesize
{\it Author Contributions:} All authors contributed to the development and writing of this paper. 
The authorship list reflects the lead authors of this paper (JHD,  TT and ENC), 
followed by two alphabetical groups. 
{\color{black} The first alphabetical group (CH and LvW) consists of authors who contributed to both the infrastructure of the KiDS data that was used in this work, and to the analysis itself.
Members of the second group are infrastructure contributors whose products are directly used in this work.}

\bibliographystyle{hapj}
\bibliography{cross_corr}

\bsp
\label{lastpage}
\end{document}